\documentclass[preprints,article,accept,moreauthors,pdftex]{Definitions/mdpi}

\firstpage{1} 
\pubvolume{1}
\issuenum{1}
\articlenumber{0}
\pubyear{2021}
\copyrightyear{2021}
\externaleditor{{Academic Editor: Masha Chernyakova}} 
\datereceived{22 July 2021} 
\dateaccepted{26 August 2021 } 
\datepublished{} 
\hreflink{https://doi.org/} 

\newcommand{\lsi}{LSI$+$61~303}
\newcommand{\hess}{HESS~J0632$+$057}
\newcommand{\psrb}{PSR~B1259$-$63}

\Title{Optical and Near-Infrared Monitoring of Gamma-Ray Binaries Hosting Be Stars}

\TitleCitation{Optical and Near-Infrared Monitoring of Gamma-Ray Binaries Hosting Be Stars}

\Author{Yuki Moritani $^{1,2,}$*\orcidA{} and Akiko Kawachi $^{3}$\orcidB{}}

\AuthorNames{Yuki Moritani and Akiko Kawachi}

\AuthorCitation{Moritani, Y.; Kawachi, A.}

\address{%
$^{1}$ \quad Kavli Institute for the Physics and Mathematics of the Universe (WPI), The University of Tokyo Institutes
for Advanced Study, 
The University of Tokyo, 5-1-5 Kashiwanoha, Kashiwa, Chiba 277-8583, Japan \\
$^{2}$ \quad Hiroshima Astrophysical Science Center, Hiroshima University, 1-3-1 Kagamiyama Higashi-Hiroshima, Hiroshima 739-8526, Japan \\
$^{3}$ \quad Department of Physics, School of Science, Tokai University, 4-1-1 Kita-kaname, Hiratsuka, Kanagawa 259-1292, Japan} 

\corres{Correspondence: yuki.moritani@ipmu.jp}

\abstract{Optical and near-infrared observations are compiled for the three gamma-ray binaries hosting Be stars: \psrb, \lsi, and \hess. 
The emissions from the Be disk are considered to vary according to the changes in its structure, some of which are caused by interactions with the compact object (e.g., tidal forces).
Due to the high eccentricity and large orbit of these systems, the interactions---and, hence the resultant observables---depend on the orbital phase. 
To explore such variations, multi-band  photometry and linear polarization were monitored for the three considered systems, using two 1.5 m-class telescopes: IRSF at the South African Astronomical Observatory and Kanata at the Higashi–Hiroshima Observatory.
}


\begin{document}
%

\section{Introduction}\label{sec1}

Gamma-ray binaries are the group of X-ray binaries 
 emitting strong, high-energy (0.1--100 GeV) and very high-energy ($>$100 GeV) gamma-rays \citep{Dubus2013}.
About 10 systems have been classified as gamma-ray binaries in the Galaxy and Large Magellanic Cloud.
The optical/near-infrared companion stars of the compact objects are either an O star or a Be star---a B-type star exhibiting emission-line profiles in Balmer lines from its circumstellar disk (Be disk) \citep{Rivinius2013}.
Companion massive stars dominate the emissions from the binaries in the optical and near-infrared bands.
In a gamma-ray binary, the relativistic particles from the compact object interact with the photon field and the low-density fast outflow (wind), emitting UV radiation from the massive star.
In the case that the compact object is the rotation-powered pulsar, when the pulsar wind collides with the stellar winds and/or the Be disk, particle acceleration is thought to occur at the shock region, where non-thermal emissions increase: this is the case considered in a pulsar-wind model.
On the other hand, in the system where mass is transferred from the stellar wind and/or the Be disk to the compact object, sufficient mass transfer leads to jet formation, such that particle acceleration occurs (microquasar model).
In either model, the stellar wind and/or the Be disk play important roles in the mass/plasma distribution, photon field, and other structures determining binary interactions.

The most fascinating fact about gamma-ray binaries is that despite many discussions (both observational and theoretical), the nature of the compact object (and, hence, the binary) has not been well established in most of the systems, except for those where a pulse has been detected (e.g., \psrb \ \citep{johnston1992} and LMC P3 \citep{Corbet2016}); however, recently, possible pulse detection has been reported for LS~5039 \citep{Yoneda2020} and \lsi \ \citep{Weng2021}.
Another interesting issue is why only a few High-Mass X-ray binaries (HMXBs) have shown such high-energy emission, among hundreds of X-ray binaries.
Furthermore, due to the diversity of the orbital period (i.e., binary separation) and the eccentricity, gamma-ray binaries provide a good experimental environment to study the mechanisms of such high-energy emissions.
As HMXBs generally have a  high eccentric orbit, the separation between the compact object and the massive star changes within each orbital cycle. 
The varying positions of the stars and the separation change the contact surface between the outflows and visibility from the observer, which causes an orbital modulation in the emissions.
For instance, light curves in X- and gamma-rays of LS~5039 reach their maximum and minimum around conjunctions.
In \psrb, outbursts occur from radio to gamma-rays around the periastron, where the pulsar approaches the Be disk.

To search for variabilities in the optical and the near-infrared bands, we monitored five gamma-ray binaries from 2010 to 2018.
The optical and near-infrared emissions of the gamma-ray binaries were dominated by those from the OB stars and the circumstellar matter.
Variations in these bands and their correlation with other wavelength bands, therefore, are expected to provide key evidence of binary interactions, as well as stellar activities contributing to the high-energy emissions.
We also monitored several HMXBs and compared their observational features with those of the gamma-ray binaries, to investigate whether any features are unique to the gamma-ray binaries.

In this paper, we report our observations of three gamma-ray binaries (\psrb, \lsi, and \hess) hosting a Be star.
If the companion is of Be type, its equatorial decretion disk adds complexity to the circumstellar environment.
The Be disk consists of a high-density, partially ionized plasma, from which the optical and near-infrared emission lines and the IR excess arise. 
Be disks in binary systems are considered to be subject to such processes as precession, warping, tidal deformation, and truncation under the influence of the compact object \citep{Rivinius2013}.
The disk may affect the pressure balance of the winds according to its density and geometrical parameters; for instance, the pulsar wind may be suppressed when it approaches the high-density region of the Be disk \citep{Torres2012}.
On the other hand, in a system with an O star, mass transfer and shock originate from the stellar winds.
As the stellar wind is symmetric, compared to the Be disk, the orientation between the two objects and the line of sight are more related to orbital variations. 
The observations of the gamma-ray binaries hosting O stars, and HMXBs will be reported in another paper, which is currently in preparation.

\psrb \ is a 48\,msec period rotation-powered radio pulsar \citep{johnston1992} in a highly eccentric ($e=$~0.87) and wide ($P_\mathrm{orb}=$ 1236.72 days) binary orbit  with a Be star,  \mbox{LS\,2883 \citep{Shannon2014}}.
The pulsed emission becomes undetectable around the periastron, and the pulsar is assumed to pass behind the Be circumstellar disk during the pulse eclipse, crossing the disk twice in each orbit: before and after periastron.
The system is the first binary to be detected with TeV gamma-rays. The radio, X-ray, GeV, and TeV flares around the periastron have been observed for multiple orbital cycles \citep{Chernyakova2020}.  
The double-peak structure of the radio, X-ray, and possibly TeV light curves are interpreted to be associated. However, the GeV flare shows a puzzling feature; it starts after the post-periastron passage, and has no counterpart in other multi-wavelengths. Discussion of the GeV emission mechanism has led to the suggestion that the infrared excess from the Be star may work as the inverse Compton target of the high-energy electrons, contributing to the optical depth in the GeV emissions \citep{vanSoelen2012}.

\lsi  \ has the smallest orbit among the three systems considered in this work; $P_\mathrm{orb}=26.496$ days and $e=0.537$ \citep{Aragona2009}.
The nature of this system has been discussed by many studies, in the frame of both the microquasar model and the pulsar-wind model.
The system is well known with both orbital modulation and super-orbital modulation (\mbox{$\sim$1700 days}) from radio to gamma-rays.
Zamanov et al. (2014) studied the correlation between H$\alpha$ line-profile parameters, V-magnitude, and $Fermi$-LAT flux.
They found anti-correlation between V-magnitude and $Fermi$-LAT flux, and correlation between Equivalent Width (EW) and $Fermi$-LAT flux \citep{Zamanov2014}.
Line-profile variability in H$\alpha$ indicates perturbations in the Be disk within one orbit, and a bright, high-speed (300--443 $\mathrm{km\;s^{-1}}$) component in the red side of the double-peaked profile before the apastron \citep{McSwain2010}.
More recently, orbital variability in its optical photometry and polarization has been studied \citep{Kravtsov2020}.
The polarization results showed a periodicity of 13.244 days---half of the orbital period.
In the $B$-, $V$- and $R$-band light curves orbital modulation with the peak shift was observed, suggesting super-orbital modulation.

\hess \ is the most puzzling system of the three considered systems, due to the uncertainty of its orbital parameters.
Since its discovery \citep{Bongiorno2011}, its orbital period has been refined by many authors, ranging from 313 to 321 days, although the proposed periods are consistent within the error.
Based on the radial velocity of the Be star, two sets of orbital parameters have been proposed.
Both show very eccentric orbit, but periastron phase are different; $e=0.83$ and $\phi_\mathrm{periastron}=0.967$ \citep{Casares2012}, or $e=0.64$ and $\phi_\mathrm{periastron}=0.663$~\citep{Moritani2018}.
Here $\phi=0$ is the the starting point of the Swift observation (JD 2454857.5) \citep{Bongiorno2011}, which is used by most of the previous studies.
In either orbit, two outbursts occur in one orbital cycle away from the periastron: A primary outburst at the phase $\phi$$\sim$0.3--0.4, with a strong peak, and a secondary outburst at $\phi$$\sim$0.6--0.9.
The X-ray and gamma-ray flux decrease to a minimum (``dip'') between the two outbursts: $\phi$$\sim$0.5.
There have been few reports regarding orbital modulation in the optical and near-infrared bands to date, but polarization variation around the periastron has been reported \citep{Yudin2017}.
Line-profile variabilities have been shown in the Balmer lines, suggesting Be disk perturbation at $\phi$$\sim$0.5--1, with the timescale of $\sim$60 and $\sim$150 days \citep{Aragona2010,Moritani2015}.
1.10-day periodic variation has been detected through TESS photometry (6000--10,000 \AA), caused by the stellar rotation \citep{Zamanov2021}.

In Sections \ref{sec:observation} and \ref{sec:results}, we describe the observations and the results, respectively.
We discuss our results in Section \ref{sec:discussions}, and give our concluding remarks in Section \ref{sec:conclusion}.

\section{Observations}\label{sec:observation}
The objective of this work is to search for variabilities in the optical and near-infrared bands in the gamma-ray binaries, which is related to the existence of the compact object. Such variabilities are often expected to have orbital modulations.
The considered targets host Be stars, but their orbital period and the X/gamma-ray activities differ widely.
We aim to compare variabilities in the optical and near-infrared bands with those in X-rays and gamma-rays to see if the variations are related to binary interactions.
Searching for commonality and difference between the three targets will provide hints to the relationship between interactions and the orbital parameters.

We monitored the three gamma-ray binaries using two telescopes, in South Africa and Japan, from 2010 and 2018.
Among the three systems reported in this work, \hess \ was visible from both telescopes, but the other systems were visible from either telescope, due to their high declination.
The observation log is summarized in Table \ref{tab:obslog}.
We also revisited our spectroscopic observations of \hess \ taken with high-dispersion spectrographs (HIDES and ESPaDOnS) \citep{Moritani2015,Moritani2018} from 2013 to 2018, to compare the photometry and linear polarization with line-profile variabilities.
We derived EW of H$\alpha$ and H$\beta$. 
Please note that part of EW(H$\alpha$) has already been published \citep{Moritani2015}.

\subsection{IRSF}

The near-infrared monitoring observations of \psrb \ and \hess \ were performed from 2010 to 2018 using the IRSF (InfraRed Survey Facility) 1.4 m telescope located at the Sutherland station of the South African Astronomical Observatory, using a three-channel infrared camera: SIRIUS (Simultaneous InfraRed Imager for Unbiased Survey) \citep{2003SPIE.4841.459N}.  SIRIUS offers $J$, $H$, and $Ks$ bands simultaneously, with a field of view of $7'.7 \times 7'.7$.

For polarimetric observations, the single-beam polarimeter, SIRPOL (SIRius POLarimetry mode) \citep{2006SPIE.6269.51K} was installed upstream of the camera. The typical accuracy of the polarization degree $\delta$ P was about 0.3\%, depending on the stability of the sky.
We observed \psrb \ around the 2010 and 2014 periastron passages for about a month in each orbital cycle, with some observations in the quiescent phase taken as reference.  For \hess , we first focused on the phase zero of the Swift observation \citep{Bongiorno2011}, which had been considered to be near the periastron \citep{Casares2012} .  After 2015, we attempted to obtain a larger coverage of the orbital phase. 

The primary data reduction was carried out using the standard pipeline software for the SIRIUS and SIRPOL. Aperture photometry was followed using the IRAF APPHOT task, with the aperture size adjusted to the FWHM of stars in the respective image. Differential photometry with multiple reference stars in the FoV was calculated, and the light curves were deduced with reference to a quiescent observation. The standard error of the zero point of instrumental magnitude was adopted for calculation of the error for each observation, including the uncertainty due to the instability of the reference stars, as well as statistical errors. 
For \hess , the target was brightest by a magnitude of about 3, compared to the stars in the FoV, which made it difficult to set an appropriate exposure time for the observation of both the target and the reference stars, as well as to reduce the errors of the light curves.
To handle this issue, we took defocused images in the latter period of the observations.
For the polarimetric observations, the average of each night was calculated. It is noted that the number of the dataset differs night by night, so the size of the error bar (standard deviation) differs accordingly. The data on the night when only one or two sets were taken have no error bar in the plotted figures. 
It is also noted that polarization was not calculated for the data in 2018.

\subsection{Kanata}

Multiple-band photometric observations of \lsi \ and \hess \ were carried out from 2012 to 2016, using the 1.5 m telescope Kanata equipped with HOWPol and HONIR at Higashi–Hiroshima Observatory in Japan.
These systems were monitored along with other HMXBs, including LS~5039, to explore orbital variabilities as well as to conduct follow-up comparisons with X-ray and gamma-ray activities.
For this purpose, the targets were observed every few days, without focusing on a certain phase.

HOWPol (Hiroshima One-shot Wide-field Polarimeter) \citep{Kawabata2008} is the optical instrument capable of imaging, polarimetry, and low-resolution spectroscopy.
Imaging mode has a 15' diameter of FoV.
Among the five filters, $VR_cI_c$ filters were used to monitor the two systems.
HOWPol was also used to monitor $R$-band linear polarization of these systems. 

HONIR (Hiroshima Optical and Near-InfraRed camera) \citep{Akitaya2014} is also capable of imaging, polarimetry and low-resolution spectroscopy, covering the optical and near-infrared bands.
The imaging mode has a 10' square of FoV.
HONIR has two arms\endnote{By design, HONIR has three arms, among which two arms were available during the observational period.}, 
each with selectable filters. 
Optical and near-infrared data were captured simultaneously.
$VRJK$ filters were used mainly, but the $H$ filter was also used on several nights at the beginning of the monitoring period.

Data reduction of the Kanata data was carried out using IRAF packages for bias/dark subtraction and flat fielding.
The same reference stars are used to measure the magnitude of \lsi \ using aperture photometry.
For \hess , there were few proper comparison stars which were close to the target and bright enough in both the optical and near-infrared bands.
Therefore, different sets of the comparison stars were used between the two instruments.
Similar to IRSF/SIRIUS photometry, differential magnitude from a given time was calculated, to compare the observations (see Appendix A.). 
For polarization data, two or three sets of images were taken, while rotating the half-wavelength plate ($0^\circ$, $22.5^\circ$, $45^\circ$ and $67.5^\circ$) and the average of these sets was calculated.
The parameters $Q$ and $U$, followed by the polarization degree and angle, were calculated from polarization data.
Typical error in polarization degree of HOWPol and HONIR is 1\% and 0.5\%,~respectively. 


$ $\begin{specialtable}[H] 
\caption{Summary of the observations.
Column 1 is the target name.
Columns 2 and 3 list the used instrument and filters, respectively.
Columns 4 and 5 list the observational period (in terms of dates and MJD) and the number of the nights during it.
\label{tab:obslog}}
\setlength{\tabcolsep}{2.6mm}
\begin{tabular}{rrccr}
\toprule
\textbf{Target}	& \textbf{Instrument}	& \textbf{Filter} & \textbf{Date}	&\textbf{\#Nights}  \\
	& 	&  & \textbf{(MJD)}	&  \\
\midrule
\psrb & IRSF/SIRIUS			& $J$,$H$,$Ks$    & 20101203--20140720 
& 38 \\
    &   &   & (55,534.0--56,858.7)  & \\
    & IRSF/SIRPOL			& $J$,$H$,$Ks$    & 20100215--20110105 & 21 \\
    &   &   & (55,535.1--55,566.1)  & \\
\midrule
\hess & IRSF/SIRIUS			& $J$,$H$,$Ks$    & 20111217--20180301 & 120 \\
    &   &   & (55,912.7--58,178.8)  & \\
    & IRSF/SIRPOL			& $J$,$H$,$Ks$    & 20101213--20180128 & 28 \\
    &   &   & (55,543.0--58,146.8)  & \\
    & Kanata/HOWPol			& $V$   & 	20121004--20150120  & 174 \\
    &   &   & (56,204.7--57,042.5)  & \\
    & Kanata/HOWPol			& $Rc$ (photo)   & 20121005--20150120  & 163 \\
    &   &   & (56,205.7--57,042.5)  & \\
    & Kanata/HOWPol			& $Rc$ (pol)   & 20121019--20150115  & 163 \\
    &   &   & (56,219.6--57,037.5)  & \\
    & Kanata/HOWPol			& $Ic$   & 20140306--20150120  & 33 \\
    &   &   & (56,722.5--57,042.5)  & \\
    & Kanata/HONIR			& $V$   & 20150501--20161102  & 15 \\
    &   &   & (57,143.5--57,694.8)  & \\
    & Kanata/HONIR			& $Rc$ (photo)   & 20150501--20161102 & 16 \\
    &   &   & (57,143.5--57,694.8)  & \\
    & Kanata/HONIR			& $Rc$ (pol)   & 20150501--20161102  & 16 \\
    &   &   & (57,143.5--57,694.8)  & \\
    & Kanata/HONIR			& $J$ (photo)  & 20141207--20161102  & 25 \\
    &   &   & (56,998.8--57,694.8)  & \\
    & Kanata/HONIR			& $J$ (pol)  & 20150422--20161102  & 16 \\
    &   &   & (57,143.5--57,694.8)  & \\
    & Kanata/HONIR			& $H$   & 	20141209--20150310  & 7 \\
    &   &   & (57,000.5--57,091.5)  & \\
    & Kanata/HONIR			& $K$   & 20141207--20161102  & 25 \\
    &   &   & (56,998.8--57,694.8)  & \\
\midrule
\lsi & Kanata/HOWPol			& $V$   & 20121004--20150120  & 90 \\
    &   &   & (56,204.5--57,042.5)  & \\
    & Kanata/HOWPol			& $Rc$ (photo) & 20121004--20150120  & 83 \\
    &   &   & (56,204.5--57,042.5)  & \\
    & Kanata/HOWPol			& $Rc$ (pol) & 20121011--20150120  & 83 \\
    &   &   & (56,211.7--57,042.5)  & \\
    & Kanata/HOWPol			& $Ic$   & 20141006--20150120    & 22 \\
    &   &   & (56,936.6--57,042.5)  & \\
    & Kanata/HONIR			& $V$   & 20150519--20161103  & 16 \\
    &   &   & (57,161.8--57,695.6)  & \\
    & Kanata/HONIR			& $Rc$ (photo)   & 20150526--20161103 & 16 \\
    &   &   & (57,168.8--57,695.6)  & \\
    & Kanata/HONIR			& $Rc$ (pol)   & 20151104--20161103 & 14 \\
    &   &   & (57,330.8--57,695.6)  & \\
    \bottomrule			
\end{tabular} \end{specialtable}			
\begin{specialtable}[H]\ContinuedFloat						
\caption{{\em Cont.}}			
\setlength{\tabcolsep}{4.7mm}
\begin{tabular}{rrccr}
\toprule
\textbf{Target}	& \textbf{Instrument}	& \textbf{Filter} & \textbf{Date}	&\textbf{\#Nights}  \\
	& 	&  & \textbf{(MJD)}	&  \\
\midrule

    & Kanata/HONIR			& $J$ (photo)  & 20141207--20161103 & 19 \\
    &   &   & (56,998.8--57,695.6)  & \\
    & Kanata/HONIR			& $J$ (pol)  & 	20150427--20161103  & 15 \\
    &   &   & (57,139.5--57,695.6)  & \\
    & Kanata/HONIR			& $H$   & 	20141207--20141220  & 3 \\
    &   &   & (56,998.8--57,011.6)  & \\
    & Kanata/HONIR			& $K$   & 20141207--20161103  & 14 \\
    &   &   & (56,998.8--57,695.6)  & \\
\bottomrule
\end{tabular}
\end{specialtable}

\section{Results}\label{sec:results}


The observed light curves and polarization showed variabilities or indications of variability.
In the following sections, the observed features of the individual systems are described.

\subsection{\psrb}

In the near-infrared band, \psrb \ showed the $\sim$0.1-mag brightening around periastron \citep{Kawachi2021}. The light curves in the {\textit J-}, {\textit H-}, and {\textit Ks-}bands were almost identical in the 2010 and 2014 orbital cycles. The brightening started no later than 10 days before the periastron and reached maximum brightness at 12--17 days after periastron. In 2014, the brightness was nearly back to the level in the quiescent phase at 70 days after periastron.
A difference in the temporal feature was observed between the light curves of {\textit J-}, {\textit H-} and {\textit Ks-}bands; thus, a characteristic track appeared on the near-infrared color--magnitude diagram. 
The brightness in the {\textit Ks-}band increased more rapidly than in the {\textit H-}band, making the infrared color ({\textit H}$-${\textit Ks}) redder at first. However, the flux increase of the {\textit H-}band gradually caught up with that of the {\textit Ks-}band, which turned blue. The infrared color became redder again as the flare decayed. 
The time lag between the bands indicated that the variation in the Be circumstellar disk first occurs in an outer region.

\begin{figure}[H]

\includegraphics[width=6.5cm]{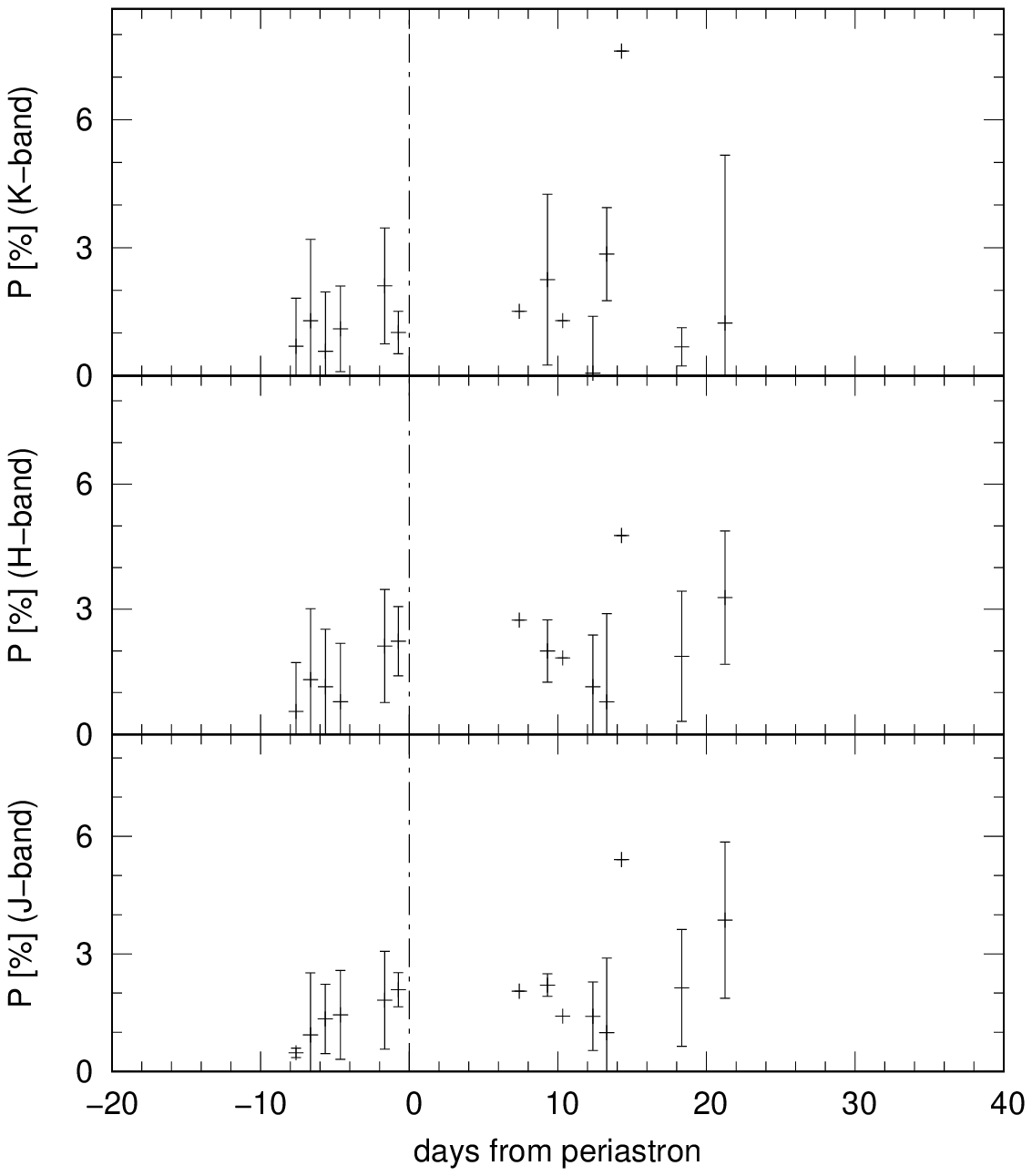}~
\includegraphics[width=6.5 cm]{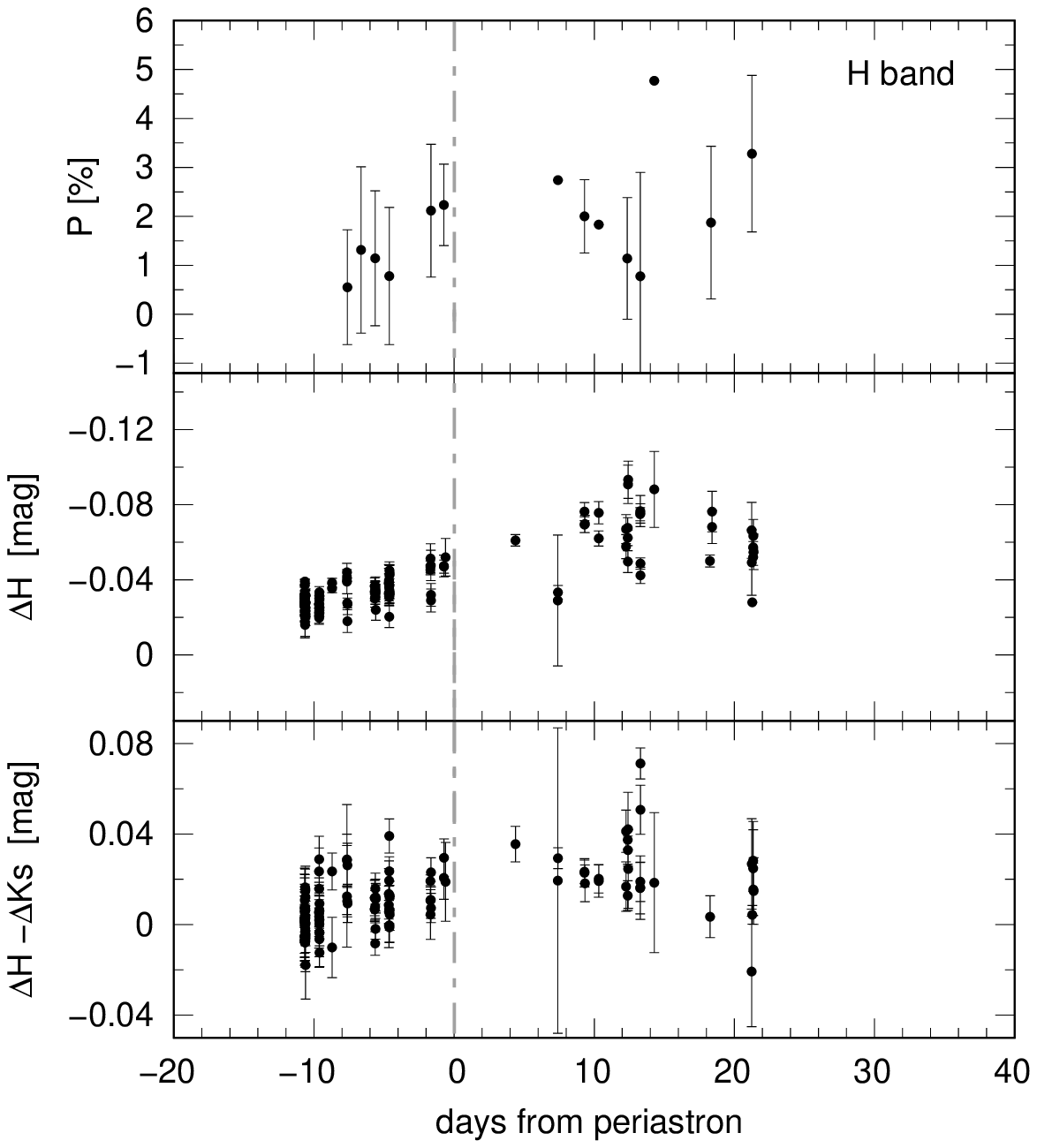}

\caption{Linear polarization of \psrb \ in the near-infrared band with IRSF/SIRPOL at the time of the 2010 periastron. 
In the left panel, the data in the $Ks$, $H$, and $J$ bands are plotted (from the top to the bottom).
The data for each night are averaged, and the error bar shows the standard deviation.
The data without the error bar indicates that only one or two sets of data were taken on that night.
In the right panel, the polarization degree in the $H$ band is compared with variations of $H$ magnitude and ($H-Ks$) color.
Please note that the data around the periastron (from $-25$ to $+40$ days) are shown in these figures.
\label{fig:pstbpol}}
\end{figure}

Figure \ref{fig:pstbpol} shows the linear polarization in the $JHKs$ bands, taken with IRSF/SIRPOL around the 2010 periastron.
The polarization degree in the near-infrared bands did not change significantly around the periastron, although it may have increased by $<$1\% in the $J$ and $H$ bands at the time of brightening.
Please note that the polarization degree observed 198 and 301 days before periastron was the same as that at periastron, within the error margins observed.
The average polarization degree in the $J$, $H$, and $Ks$ bands were $1.4 \pm 0.6$\%, $1.5 \pm 0.7$\% and $1.3 \pm 0.8$\%, respectively.
Johnston et al. (1996) have reported that the linear polarization degree in the radio bands was very high (several tens of \%), but depolarization was observed around the periastron, when the pulsar was thought to be behind the Be disk \citep{Johnston1996}.

\subsection{\lsi}

\begin{figure}[H]
\begin{tabular}{cc}
\includegraphics[width=6.5 cm]{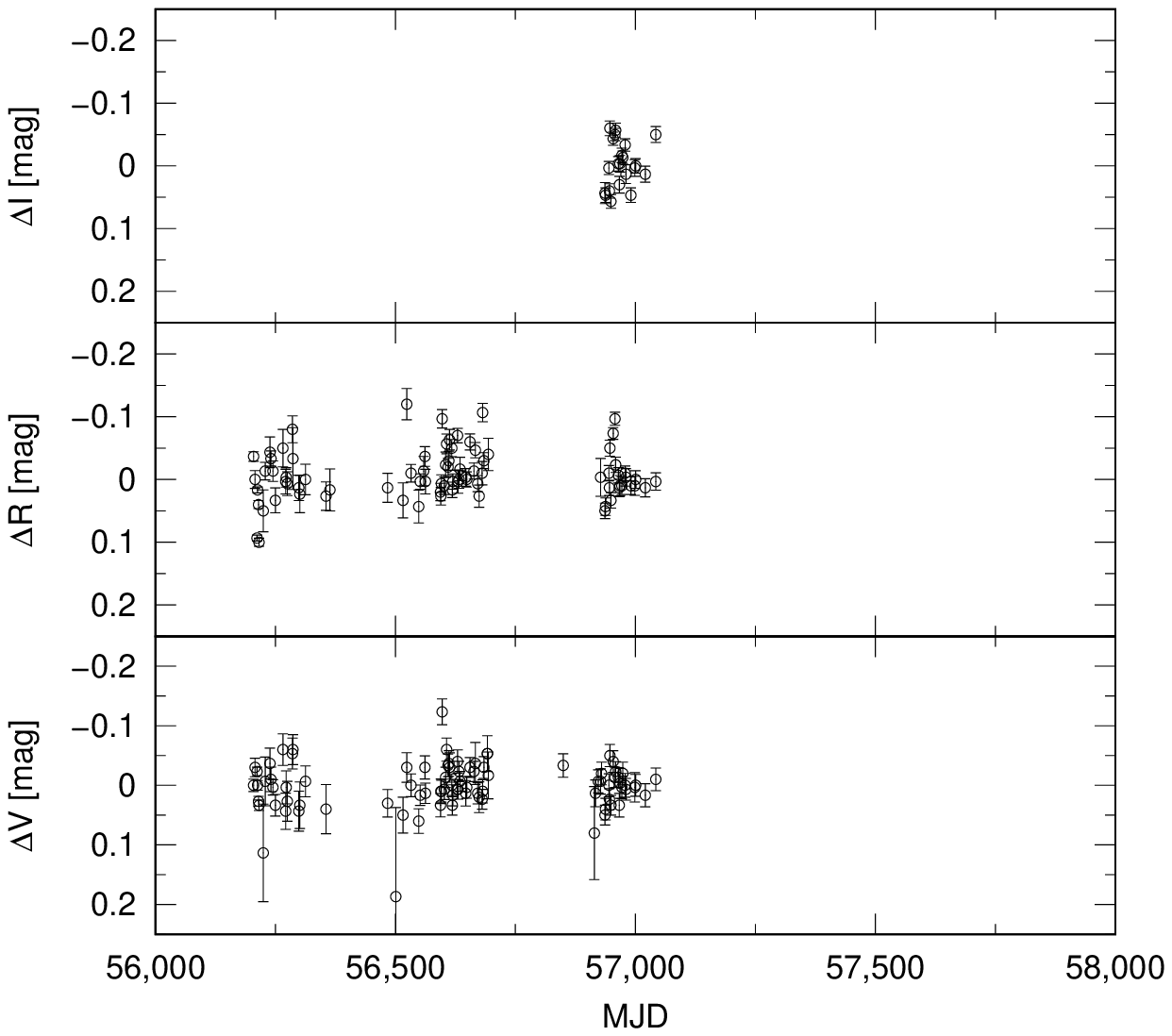}
& 
\includegraphics[width=6.5 cm]{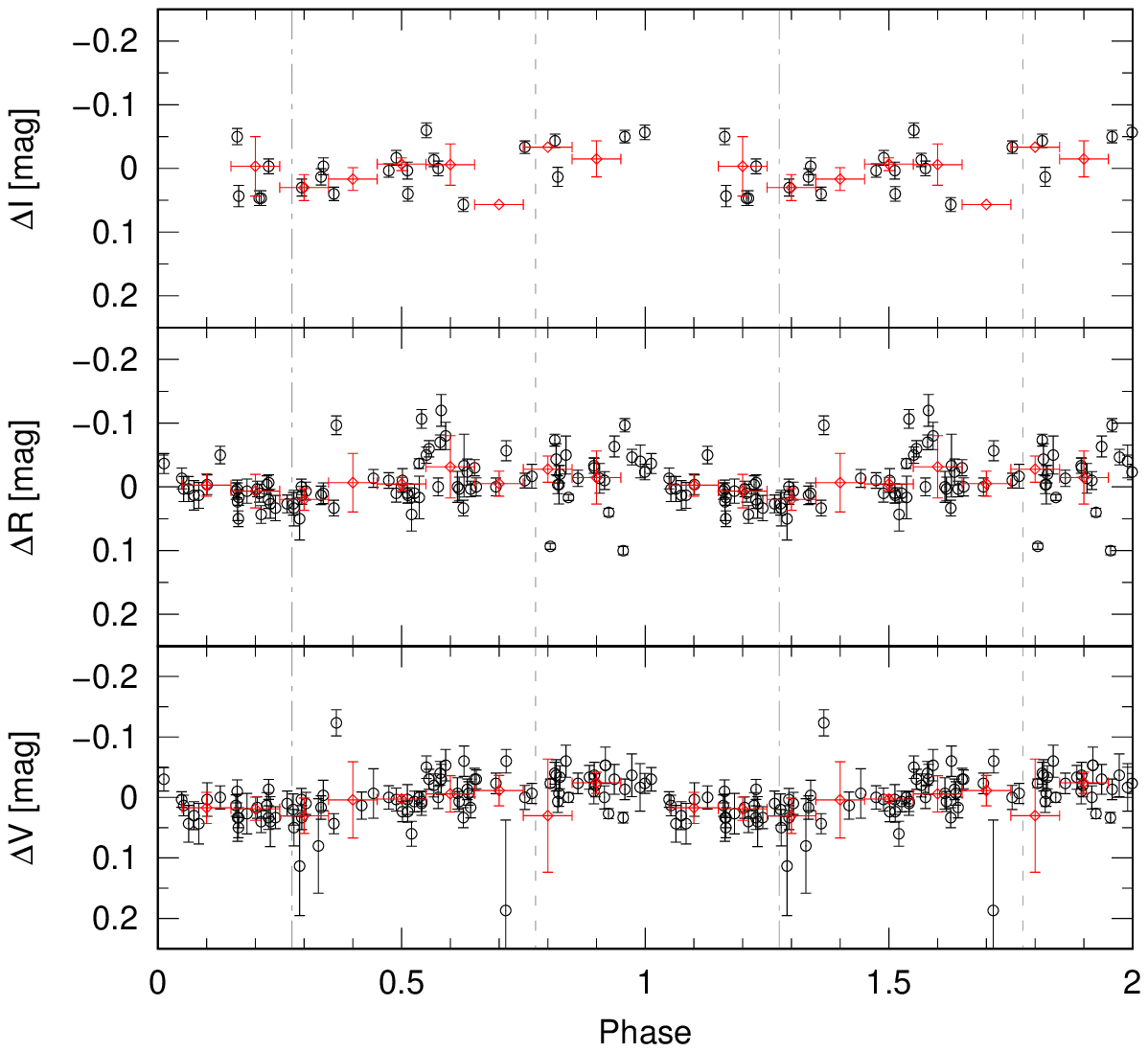}
\end{tabular}
\caption{Optical photometry of \lsi \ with Kanata/HOWPol.
From top to bottom, differential magnitude in the $I$, $R$, and $V$ bands are plotted.
The left and right panels show the variation over the entire observational period and that folded with the orbital period \citep{Aragona2009}.
In the right panel, the data in a bin of 0.1 of phase are overplotted with red points.
The data are plotted for two cycles in the right panel.
The vertical dash-dotted and dotted lines indicate the periastron and the apastron.
\label{fig:lsiPhoto_howpol}}
\end{figure}

\begin{figure}[H]
\begin{tabular}{cc}
\includegraphics[width=6.5 cm]{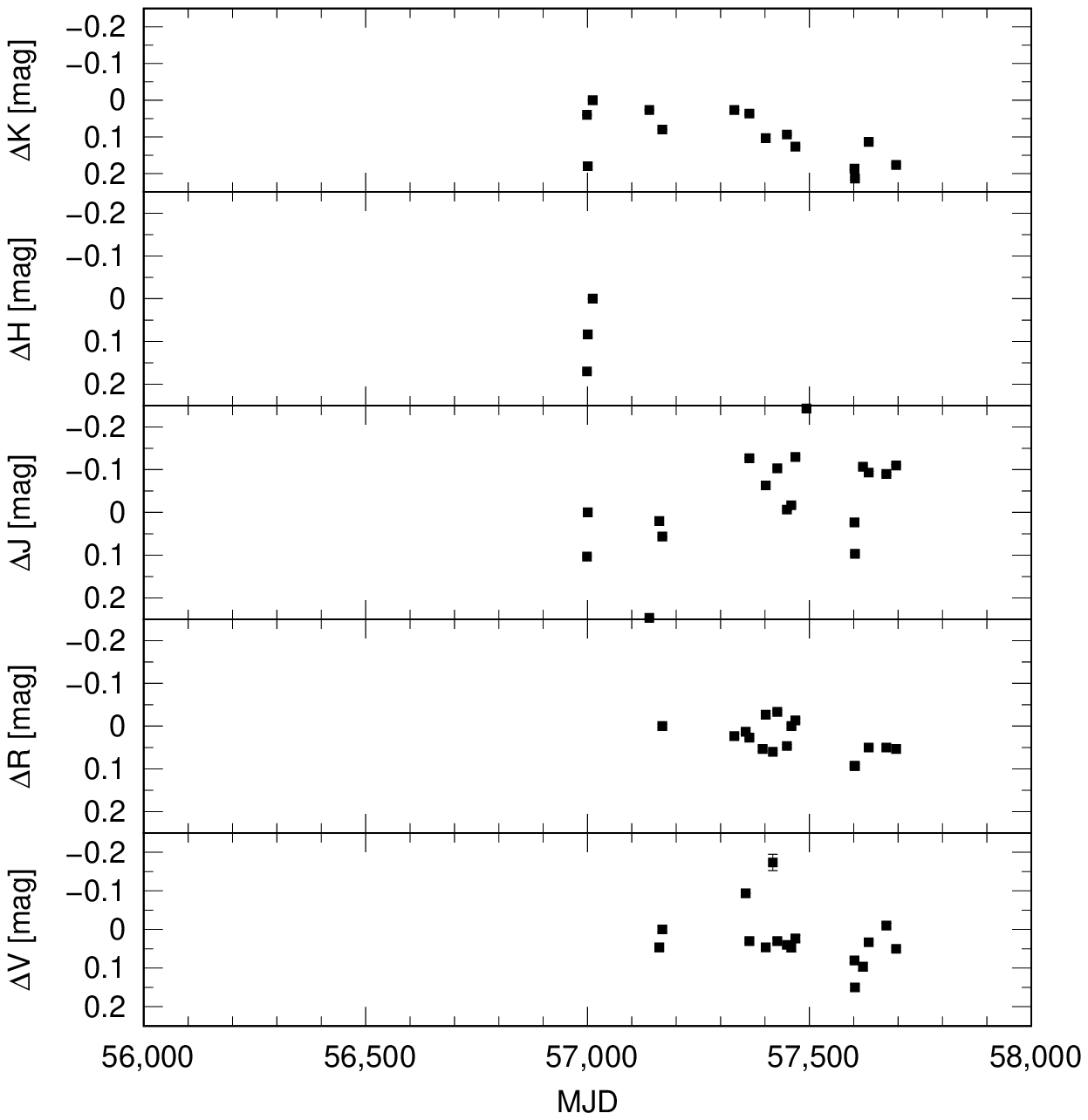}
& 
\includegraphics[width=6.5 cm]{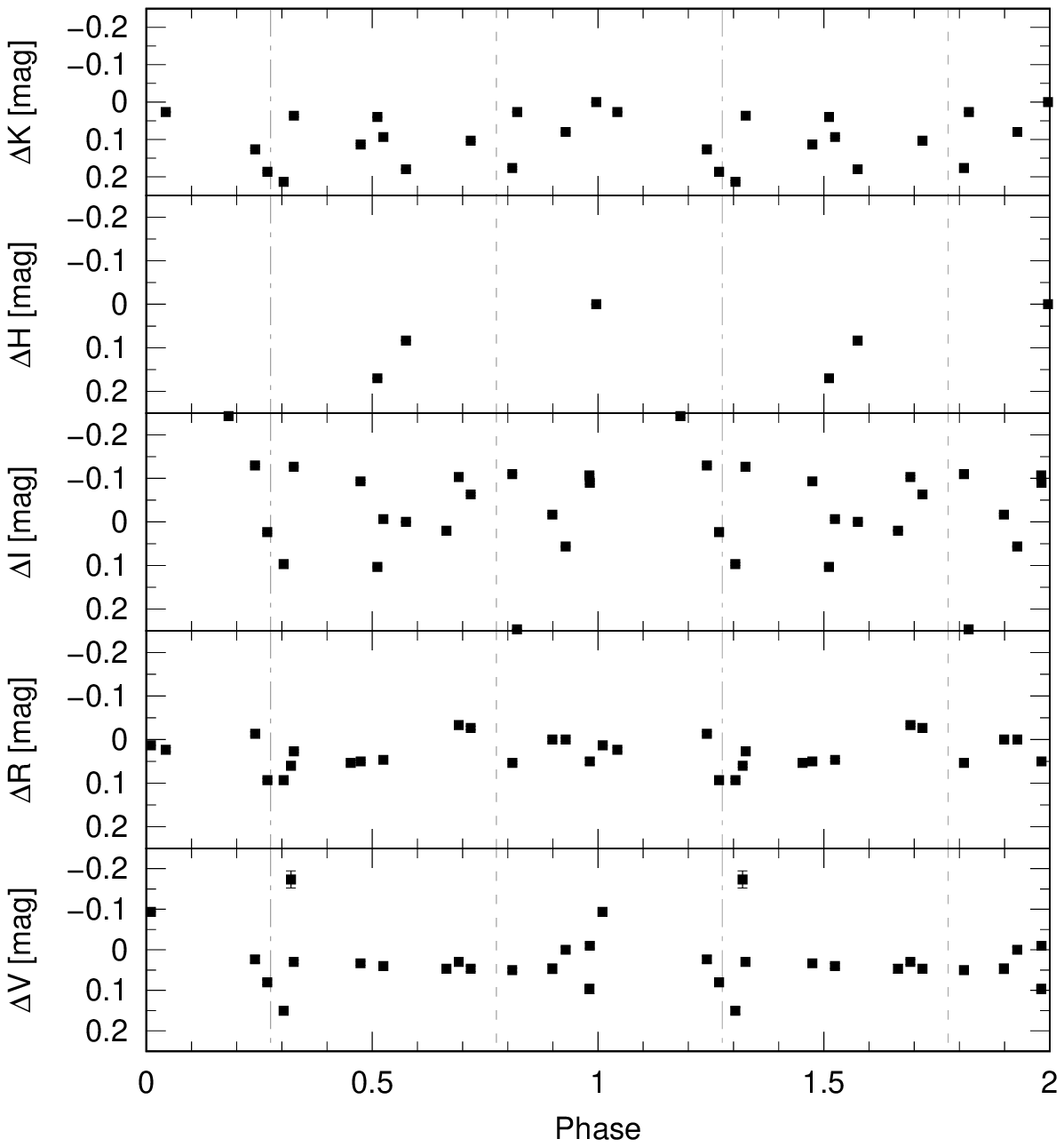}
\end{tabular}
\caption{Optical and near-infrared photometry of \lsi \ with Kanata/HONIR in the whole observation period (left) and regarding the phase (right).
From top to bottom, differential magnitude in the $Ks$, $H$, $J$, $R$, and $V$ bands are plotted. The vertical dash-dotted line and dashed line indicate periastron and apastron, respectively.
\label{fig:lsiPhoto_honir}}

\end{figure}

Figures \ref{fig:lsiPhoto_howpol} and \ref{fig:lsiPhoto_honir} show the photometric variation of \lsi \ in the optical and near-infrared bands.
Orbital modulation is clearly seen in the $V$ and $R$ bands by HOWPol; the brightness increases from around  periastron to apastron, then decreases from apastron to periastron.
The binned light curve (shown in red) showed the amplitude of the variation is $\sim$0.05 mag, which is consistent with the previous study \citep{Kravtsov2020}.
Although the phase coverage is not sufficient, the same sinusoidal feature is seen in the $I$ band light curve.
HONIR data showed larger scattering than HOWPol data. In particular, 
high scattering in the near-infrared data made it difficult to discuss the associated variations.
This may be caused by stellar activities on shorter/larger time scales, but the lack of data made it impossible to reach any conclusion.
This sinusoidal variation of HOWPol light curve agreed with previous studies (\citep{Kravtsov2020,Zamanov2014} for instance).
Both HOWPol and HONIR light curves suggest a short, $\sim$0.1--0.2-mag brightening a few days after periastron {\mbox{($\phi \sim$ 0.3--0.35)}} in the optical bands.

\begin{figure}[H]
\begin{tabular}{cc}
\includegraphics[width=6.5 cm]{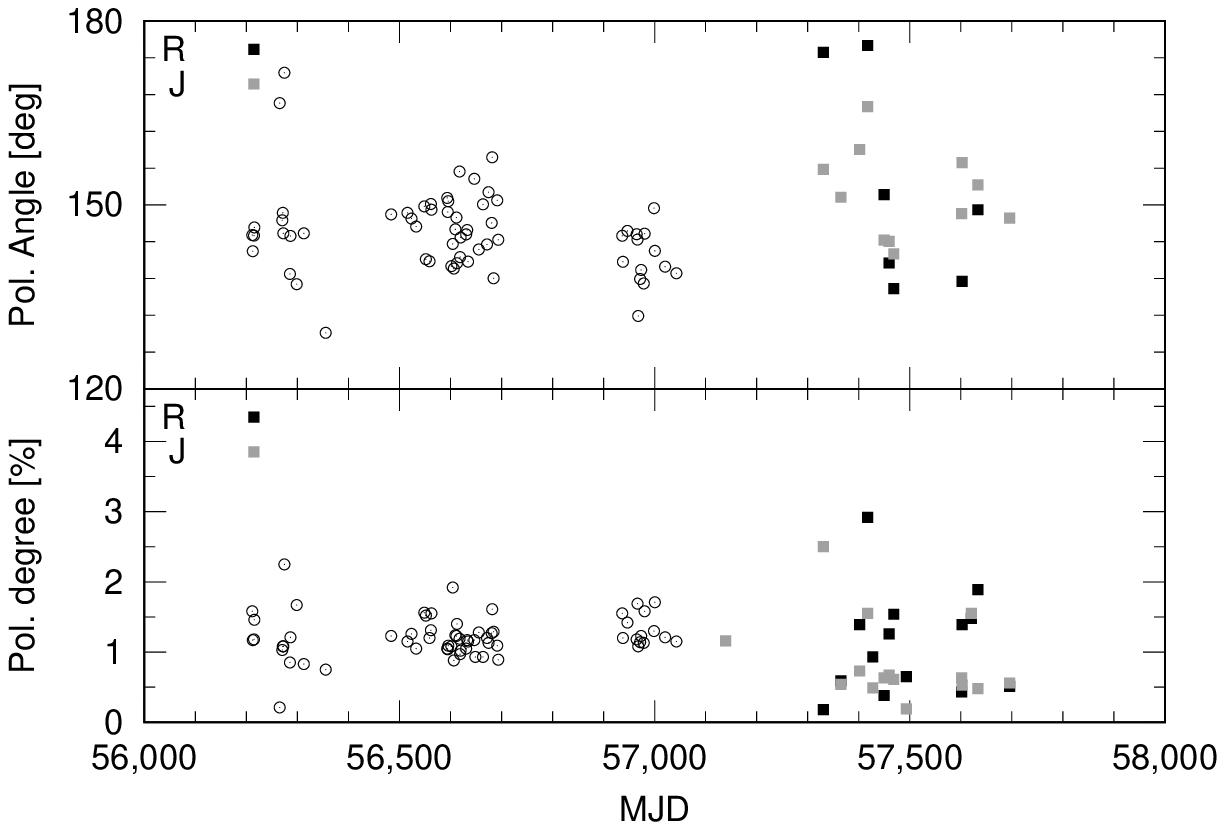}
& 
\includegraphics[width=6.5 cm]{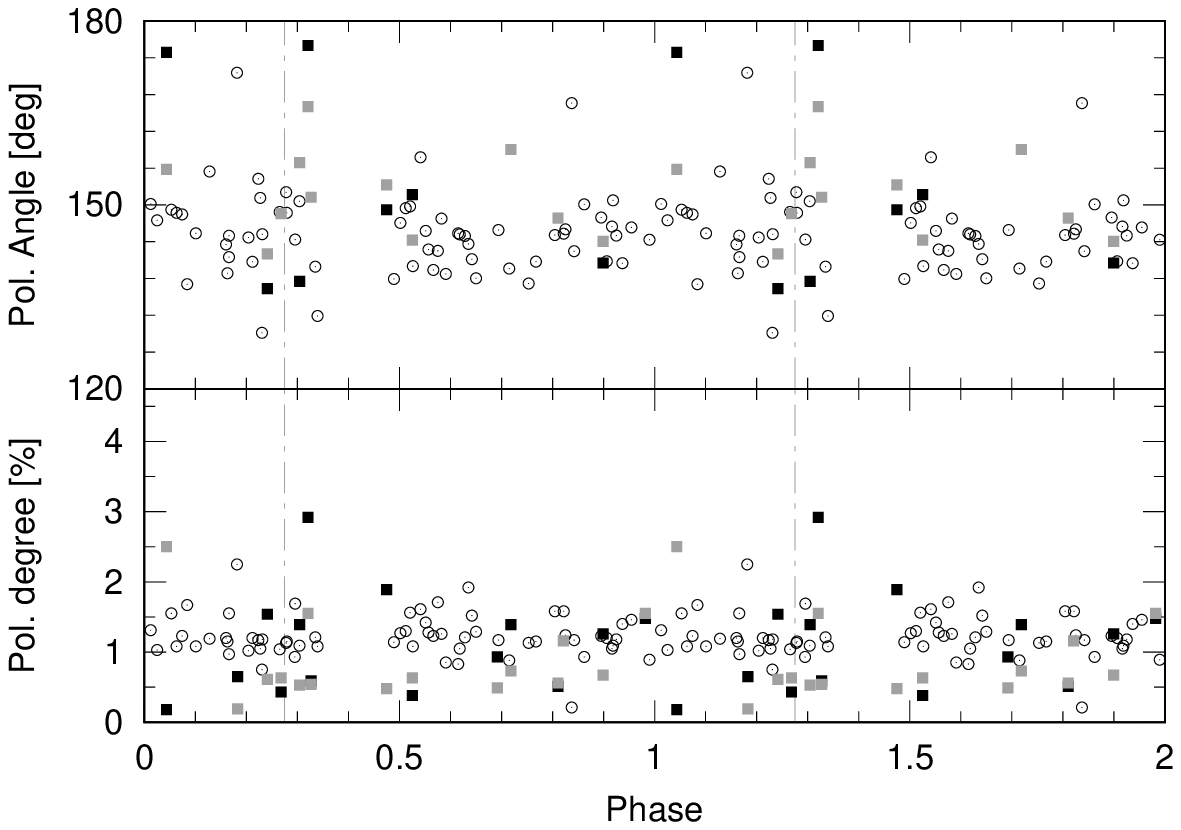}
\end{tabular}
\caption{$R$- (black) and $J$-band (gray) polarization data of \lsi \ with Kanata/HOWPol (open circles, until MJD 57037, $R$-band) and Kanata/HONIR (filled square, from MJD 57143, $R$- and $J$-bands), over the whole observation period (left) and regarding the phase (right).
\label{fig:lsiPol}}
\end{figure}

Figure \ref{fig:lsiPol} shows the polarimetric variation in the $R$ and $J$ bands.
The polarization degree ($\sim$1\%) is very small, comparable to the order of the error of the instruments.
No orbital modulation was suggested.
Periodic variability has been reported to have a period of 13.244 days---half of the orbital period \citep{Kravtsov2020}.
The average polarization degree and angle were $1.2 \pm 0.2 $\% and $144.4 7.2$ degree in the $R$ band, and $0.6 \pm 0.2$\% and $151.7 \pm 7.3$ degree in the $J$ band.
The result for the $R$ band was consistent with that of previous studies; $1.25 \pm 0.04$\% and $135.4 \pm 1.0$ degree \citep{Kravtsov2020}.

\vspace{-16pt}

\subsection{\hess}

\begin{figure}[H]
\widefigure

\includegraphics[width=6.0 cm]{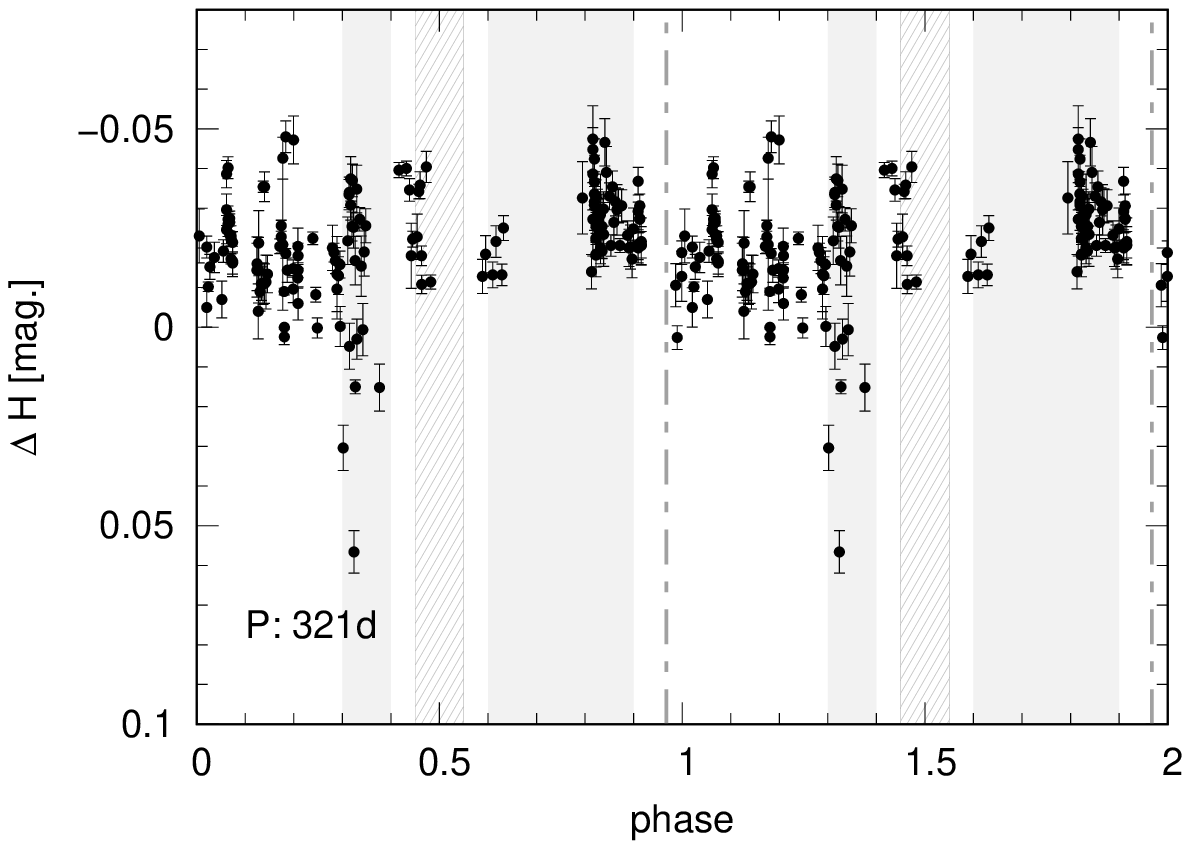}~
\includegraphics[width=6.0 cm]{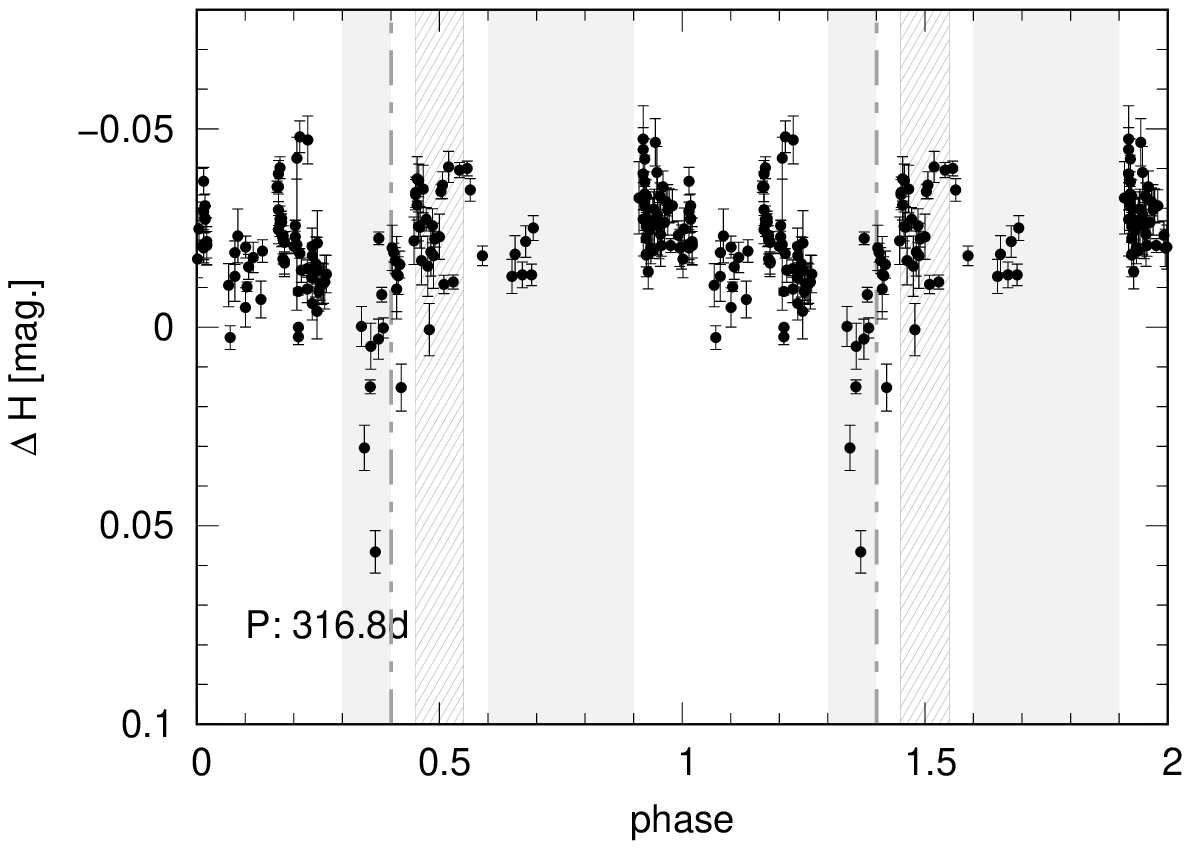}~
\includegraphics[width=6.0 cm]{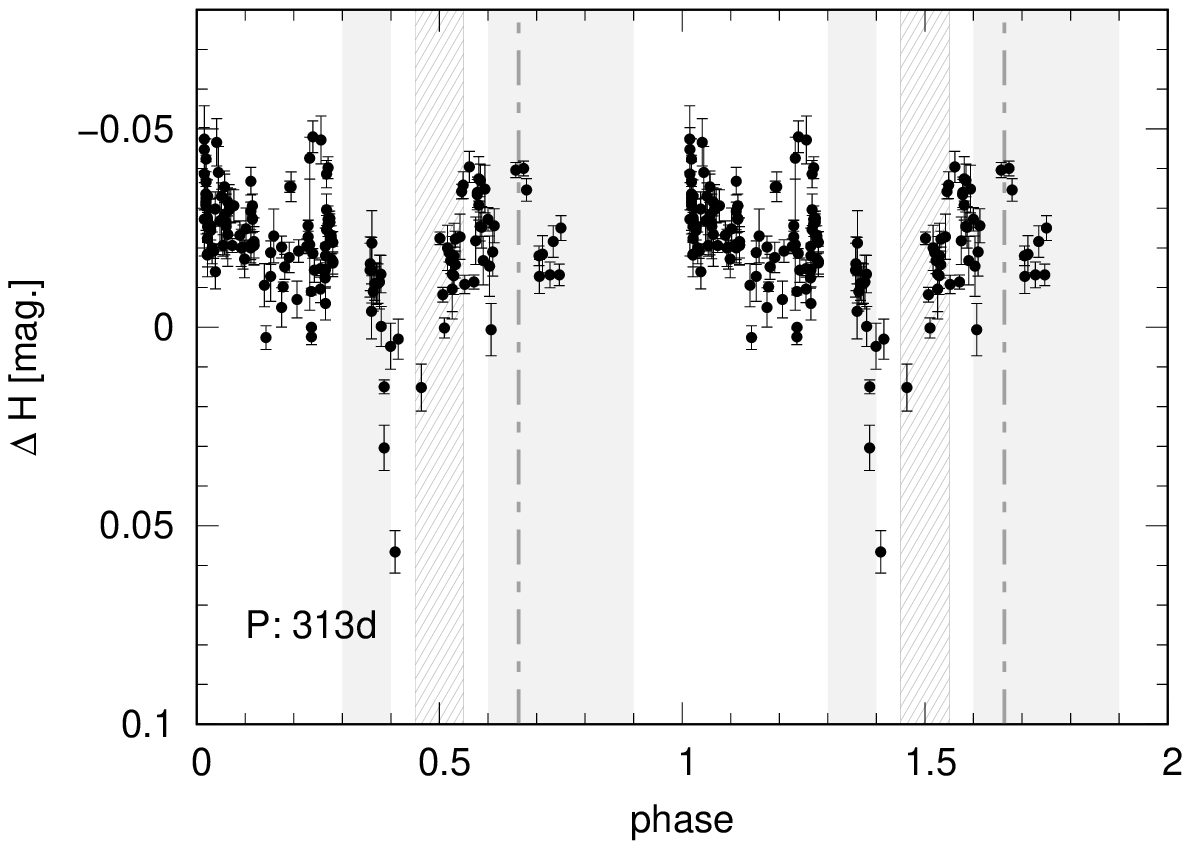}

\caption{$H$-band relative magnitude of \hess \ by IRSF/SIRUIS folded with three different orbital periods: 321~days~\citep{Bongiorno2011}  (left 
), 316.8 days \citep{Malyshev2019} (middle) and 313 days \citep{Moritani2018} (right).
The phases of the X-ray outbursts ($\phi$$\sim$0.3--0.4 and $\sim$0.6--0.9 ) are marked with the light-gray shaded area, and that of the dip ($\phi$$\sim$0.45--0.55) with the striped area. The vertical dash-dotted line in each figure indicates the periastron phase offered by the same reference of the folding~period.
\label{fig:hessPhotoIRSF_multip}}
\end{figure}

Figure \ref{fig:hessPhotoIRSF_multip} displays the $H$-band magnitude folded with three orbital periods proposed for \hess \ with which the orbital parameters has been calculated: \mbox{321 days \citep{Bongiorno2011}}, 316.8 days \citep{Malyshev2019}, and 313 days \citep{Moritani2018}. As described in Section \ref{sec1},  the orbital parameters of \hess \ still have large uncertainties.
We have selected the orbital periods which represent the range of proposed period to check the effect of its uncertainty. 
The flux decreases around $\phi$$\sim$0.3--0.4, where the primary X-ray and gamma-ray outbursts occur, though the variation is not sufficiently significant considering the error bars and the scarce data points.
  Our data do not offer enough sensitivity to determine the orbital period, but, as 
seen in the figures, the several-day difference in the orbital period led to a difference in the phase distribution of the observed data, at  $\phi$ of  0.8--1.3 in particular, when the data over several orbital cycles are combined.
Hereafter, we use $P_\mathrm{orb}=313$ days which was deduced by the radial velocity of the observations with a wider coverage of the orbit to discuss the orbital variability.

Figure \ref{fig:hessPhotoIRSF} shows the $J$-, $H$-, and $Ks$-band photometry data, compared with the Equivalent Width (EW) of H$\alpha$ and H$\beta$ lines.
A decrease of magnitude is also seen in the $J$-band.
The EW of H$\alpha$ did not show a clear orbital modulation.
The absolute value of EW of H$\beta$, on the other hand, showed a hint of modulation with two minima in one orbital cycle; $\phi$$\sim$0.3 and $\phi$$\sim$0.9. 
The phase at its minimum is different from those of  the $J$ and $H$ flux ($\phi$$\sim$0.3--0.4).
Figure \ref{fig:hessPhotoIRSF_h_jh} compares the color change ($\Delta J-\Delta H$) with respect to the variation in the $J$ band ($\Delta J$).
The color--magnitude diagram suggests that when the flux decreases at $\phi$$\sim$0.3--0.4, the system becomes redder.
In contrast to \psrb \ \citep{Kawachi2021}, it seems that no color transition was observed while the magnitude was increasing and~decreasing.
\end{paracol}
\begin{paracol}{2}
\switchcolumn
\vspace{-30pt}

\begin{figure}[H]
\includegraphics[width=7.0 cm]{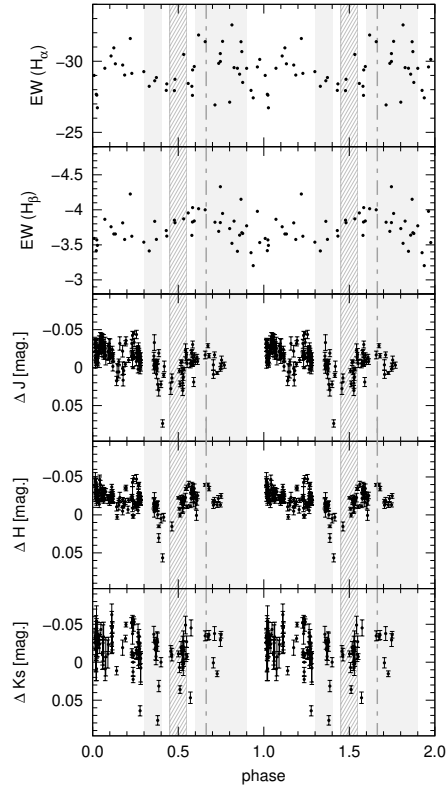}
\caption{Relative magnitude in the near-infrared band of \hess \ (IRSF/SIRUIS), compared with EW of H$\alpha$ and H$\beta$ \citep{Moritani2015,Moritani2018}.
From the top to bottom, EW(H$\alpha$), EW(H$\beta$), $J$, $H$, and $Ks$ magnitude are plotted.
The orbital period (313 days) to fold the data, and the phase at periastron (\mbox{$\phi_\mathrm{periastron}=0.663$}) are cited from \citet{Moritani2018}.
See Figure \ref{fig:hessPhotoIRSF_multip} for the description of the gray and striped~area.
\label{fig:hessPhotoIRSF}}
\end{figure}   

\begin{figure}[H]
\includegraphics[width=6.5 cm]{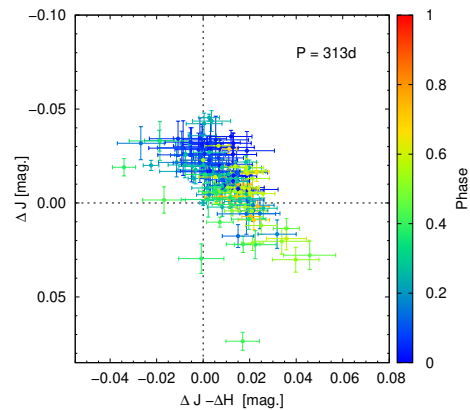}
\caption{Color ($\Delta J - \Delta H$)--magnitude ($\Delta J$) diagram for \hess \ by IRSF/SIRUIS.
The color indicates the phase $\phi$, calculated for the orbital period of 313 days.
\label{fig:hessPhotoIRSF_h_jh}}
\end{figure}

Figures \ref{fig:hessPhoto_howpol} and \ref{fig:hessPhoto_honir} show the light curves in the optical and near-infrared bands taken by Kanata/HOWPol and Kanata/HONIR.
As shown in the IRSF light curves, HONIR data showed a decrease of $V$ band magnitude around a phase of 0.25, which looks earlier than the near-infrared band (Figure \ref{fig:hessPhotoIRSF_multip}, right).
\begin{figure}[H]
\begin{tabular}{cc}
\includegraphics[width=6.5 cm]{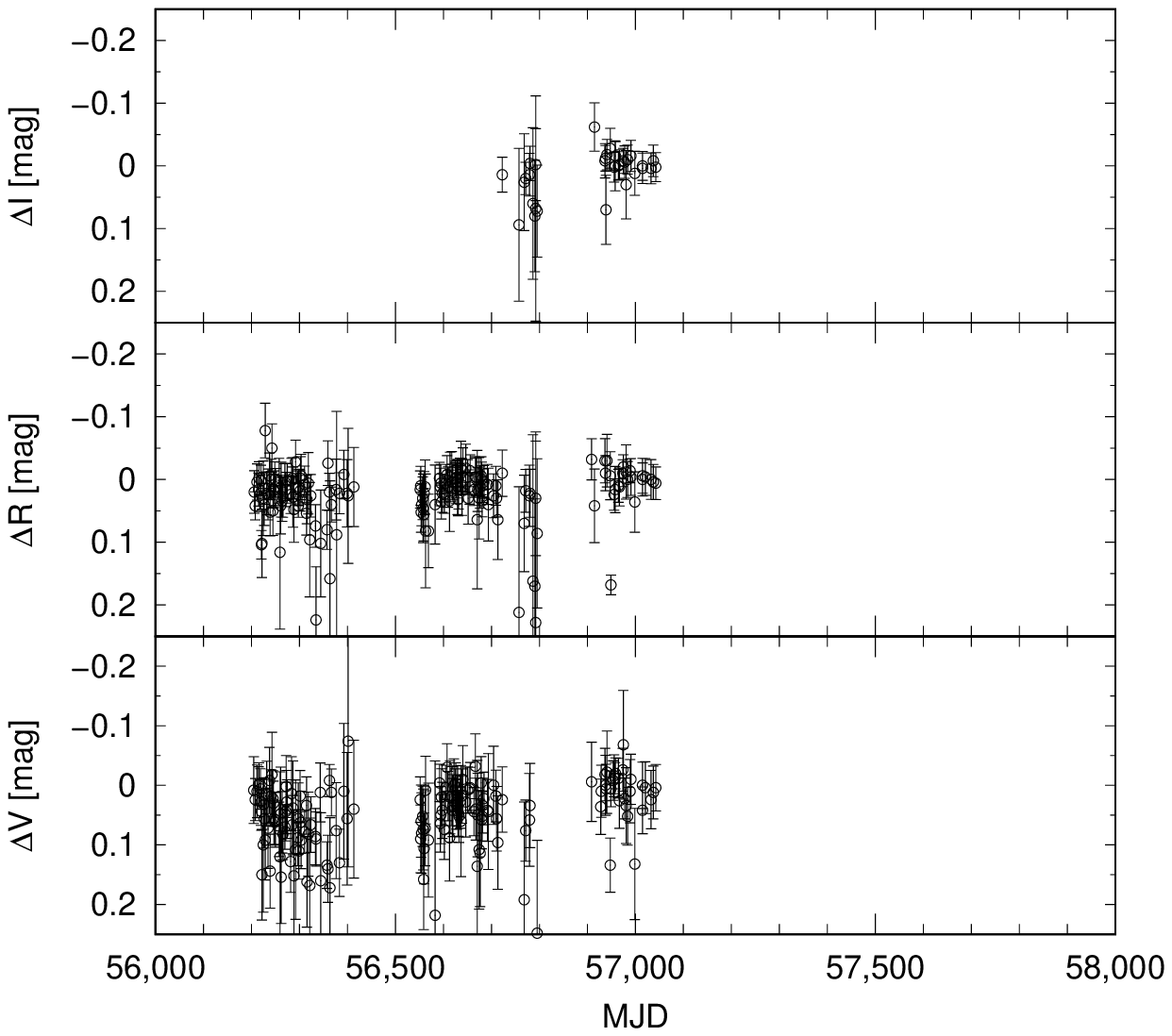}
& 
\includegraphics[width=6.5 cm]{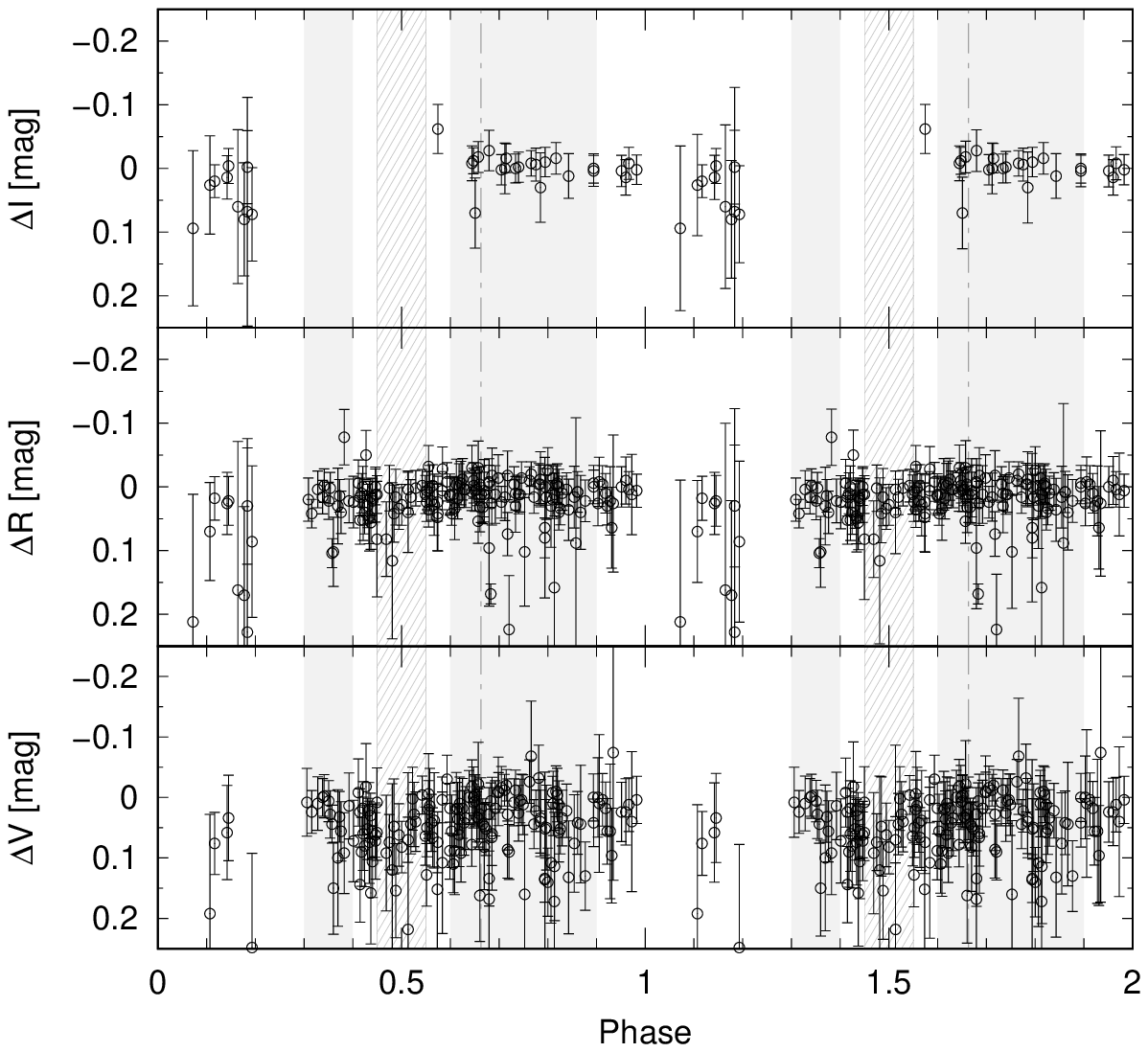}
\end{tabular}
\caption{The same as Figure \ref{fig:lsiPhoto_howpol}, but for \hess  \ by Kanata/HOWPol.
See Figure \ref{fig:hessPhotoIRSF_multip} for the description of the gray and striped area.
\label{fig:hessPhoto_howpol}}
\end{figure}

Figure \ref{fig:hessPolRSF} shows the result of linear polarization in the $J$-, $H$-, and $Ks$-bands observed with IRSF, while
Figure \ref{fig:hessPol} shows the result of linear polarization in the $R$- and $J$-bands observed with Kanata.
The averaged value of the polarization degree observed with IRSF/SIRPOL was $2.7 \pm 0.5$\% ($J$), $2.1 \pm 0.8$\% ($H$) and $2.3 \pm 0.5$\% ($Ks$), and that with Kanata/HOWPol+HONIR is $2.6 \pm 0.4$\% ($J$) and $3.8 \pm 0.2$\% ($R$).
The polarization degree in the J-band was the same between IRSF/SIRPOL in 2011--2011 and Kanata/HONIR in 2014--2016. 
Polarization in the near-infrared bands was smaller than that in the $R$ band, but the trend appeared similar between them; the polarization degree in both of the $J$- and $R$-bands decreased and increased again around MJD 57,400. 
Over several years of the observation period, no significant change was observed in linear polarization.
This agreed with the fact that the H$\alpha$ profiles did not change their intensity drastically within the observational periods \citep{Casares2012,Moritani2015,Moritani2018}.

The feature of orbital modulation is slightly seen in the $R$ band; the polarization degree increased by 0.5\% when the phase changed from 0.5 to 0.8 and then decreased, although the amplitude was comparable to measurement error.
The polarization angle did not change clearly.
Yudin et al. (2017) have reported multicolor polarimetry in the optical band at different orbital phases \citep{Yudin2017}.
This work shows a result consistent with theirs: the polarization degree and angle in the $R$-band were $3.83\pm 0.15$\%, $171.3 \pm 3$ degree (JD 2,457,369) and $3.74\pm 0.15$\%, $167.9 \pm 3$ degree (JD 2,457,369). 
Additionally, note that intra-night variation in the polarization degree has also been suggested, with the amplitude of $\sim$0.5\% (in the $V$ band) \citep{Yudin2014}.



\vspace{-12pt}

\begin{figure}[H]
\begin{tabular}{cc}
\includegraphics[width=6.5 cm]{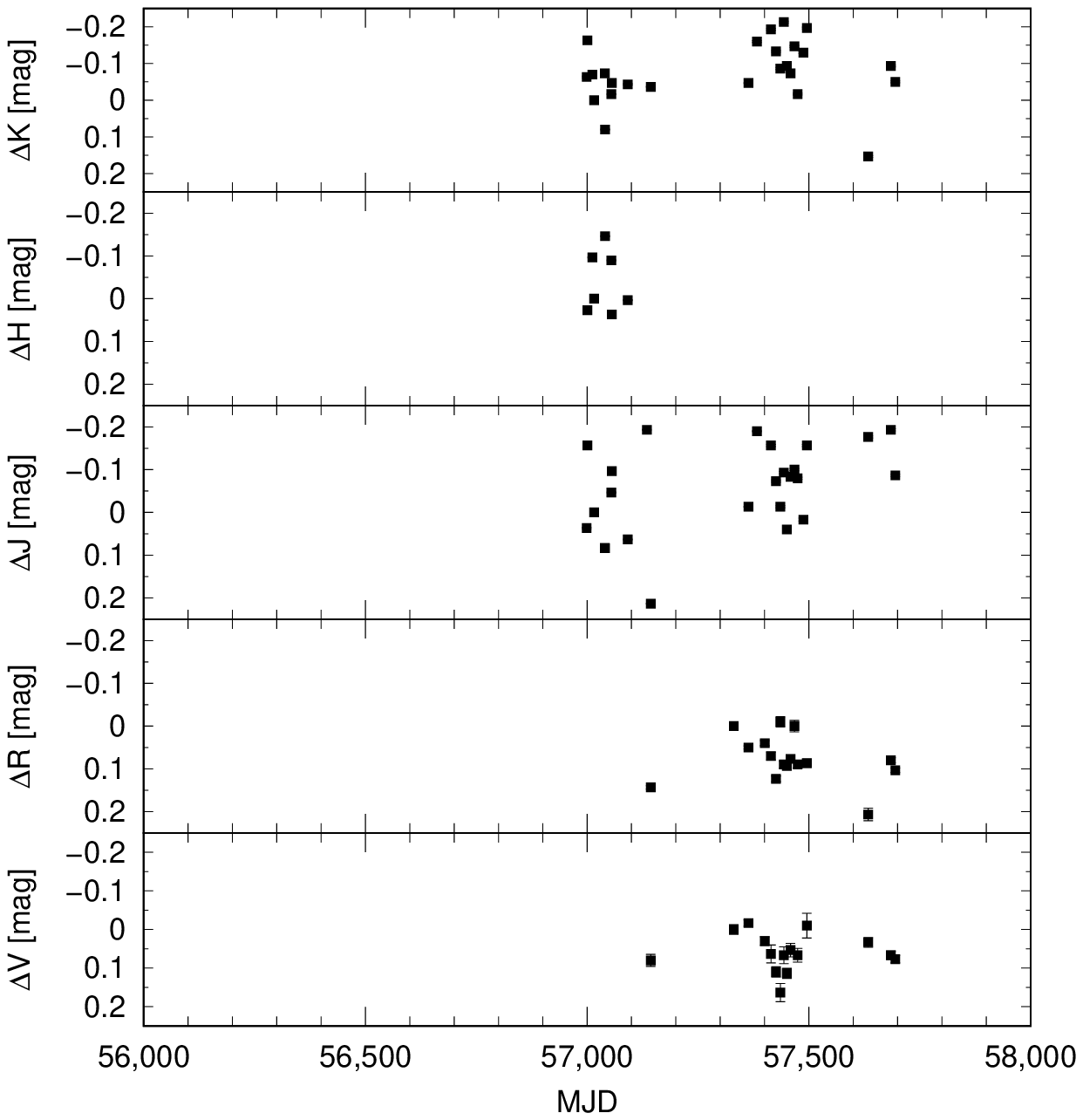}
& 
\includegraphics[width=6.5 cm]{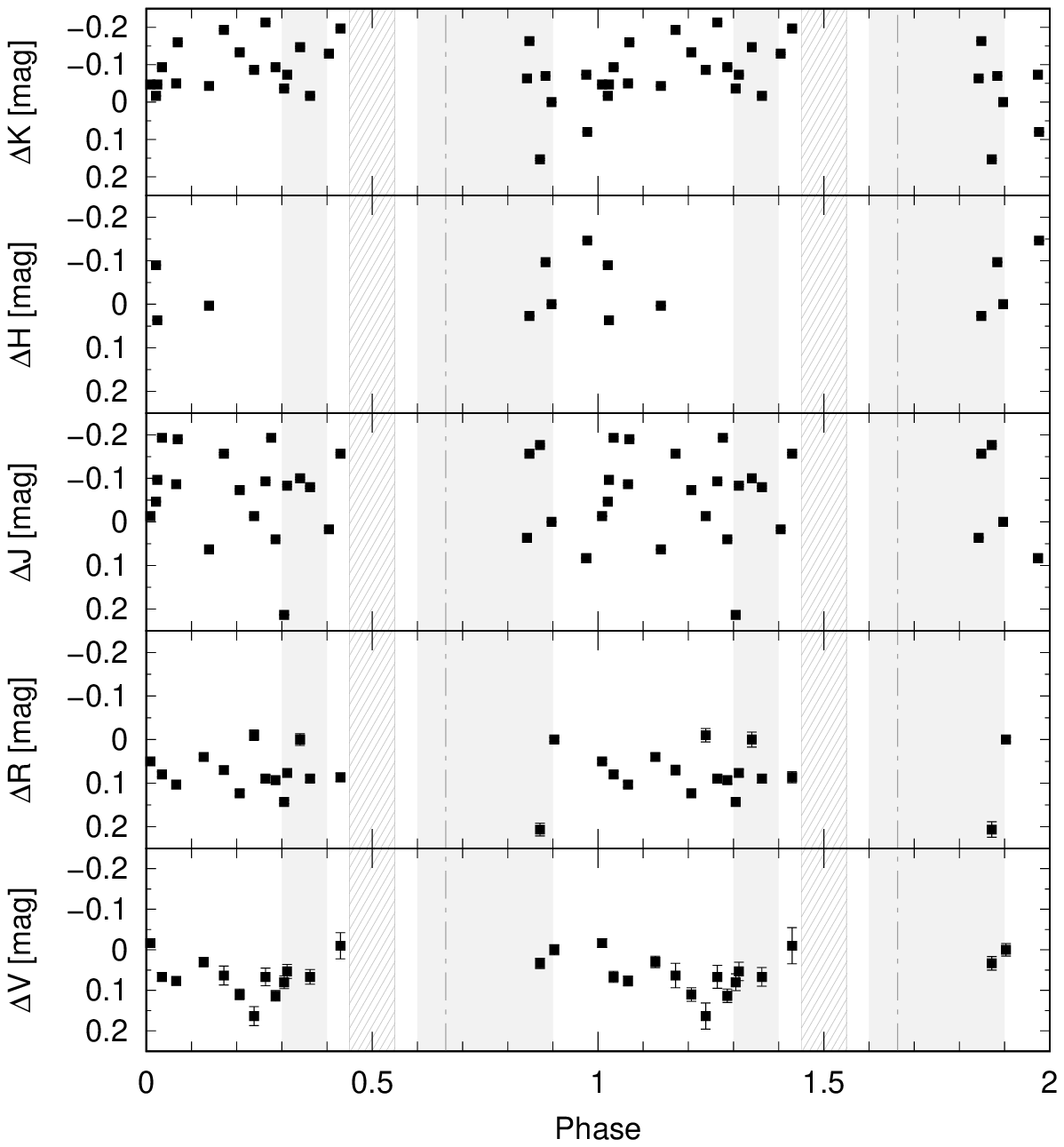}
\end{tabular}
\caption{The same as Figure \ref{fig:lsiPhoto_honir}, but for \hess  \ with Kanata/HONIR.
See Figure \ref{fig:hessPhotoIRSF_multip} for the shaded and striped area.
\label{fig:hessPhoto_honir}}
\end{figure}   
\vspace{-10pt}

\begin{figure}[H]
\begin{tabular}{cc}
\includegraphics[width=6.5 cm]{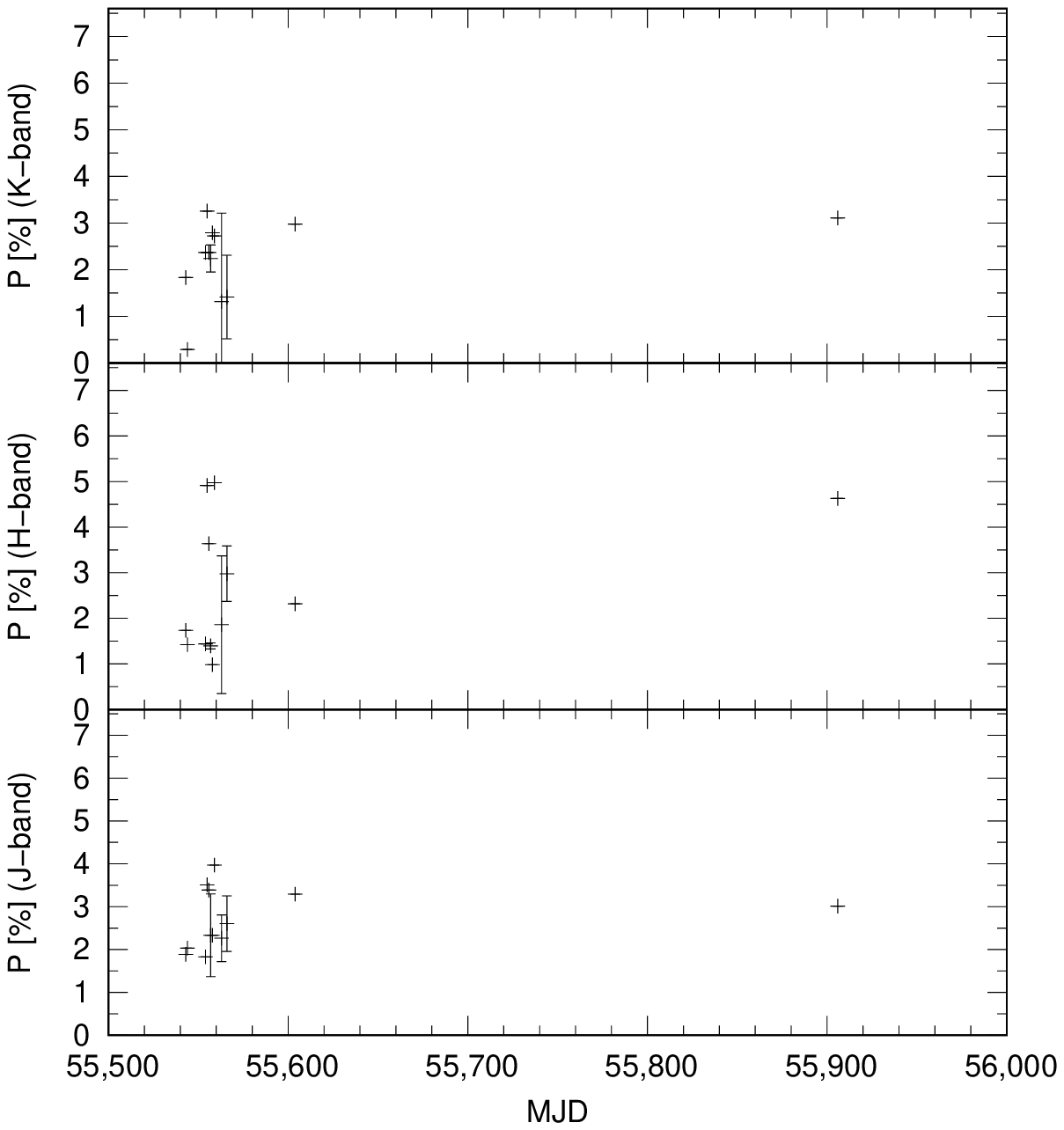}
& 
\includegraphics[width=6.5 cm]{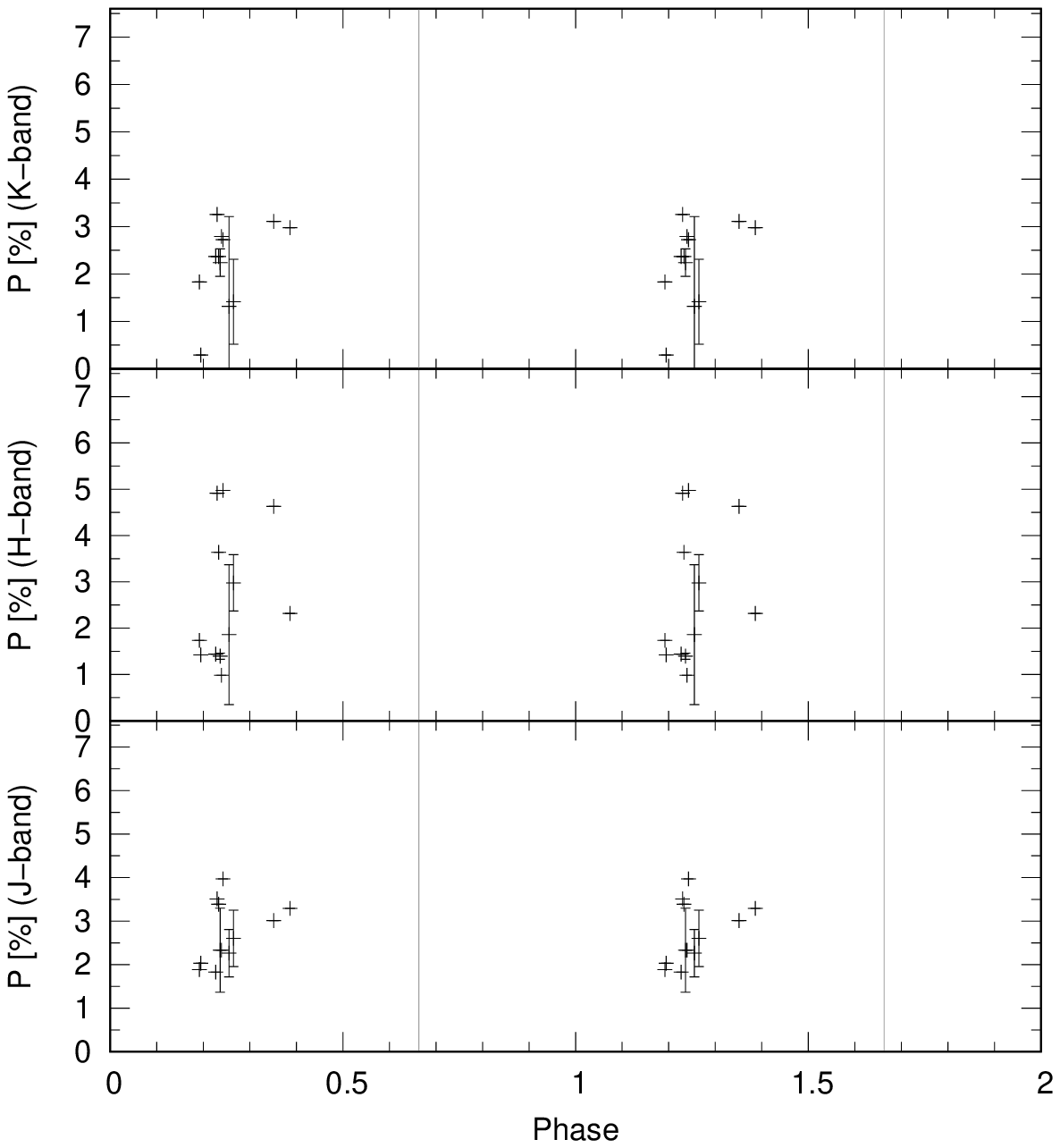}
\end{tabular}
\caption{Liner polarization of \hess \ in the $J$-, $H$-, and $Ks$-bands with IRSF/SIRPOL.
From the top to the bottom, Polarization degree in the $K$-, $H$-, and $J$-bands are plotted.
The left and right panels show variation in time (MJD) and the orbital phase for $P_\mathrm{orb}=313$ days, respectively.
The two cycles are plotted in the right panel, and the vertical dash-dotted line indicates the periastron~\citep{Moritani2018}.
\label{fig:hessPolRSF}}
\end{figure}   

\begin{figure}[H]
\begin{tabular}{cc}
\includegraphics[width=6.5 cm]{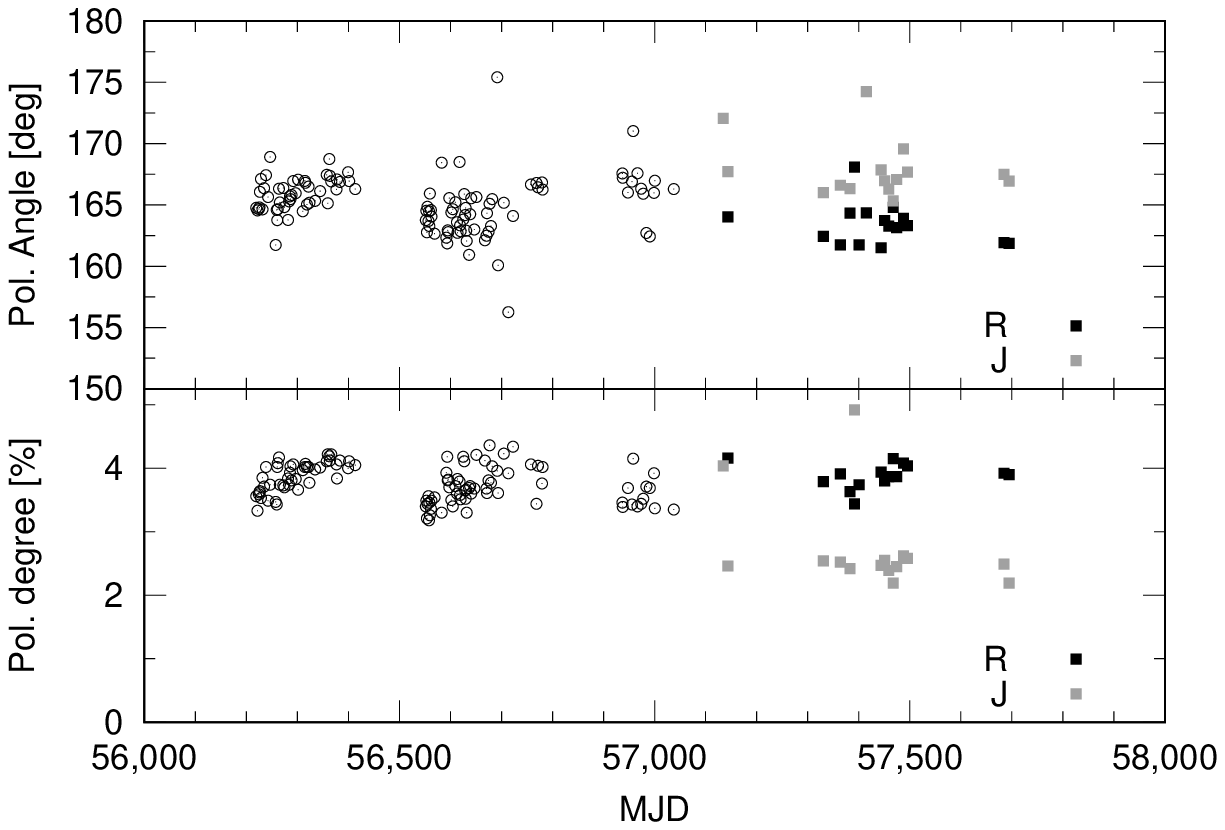}
& 
\includegraphics[width=6.5 cm]{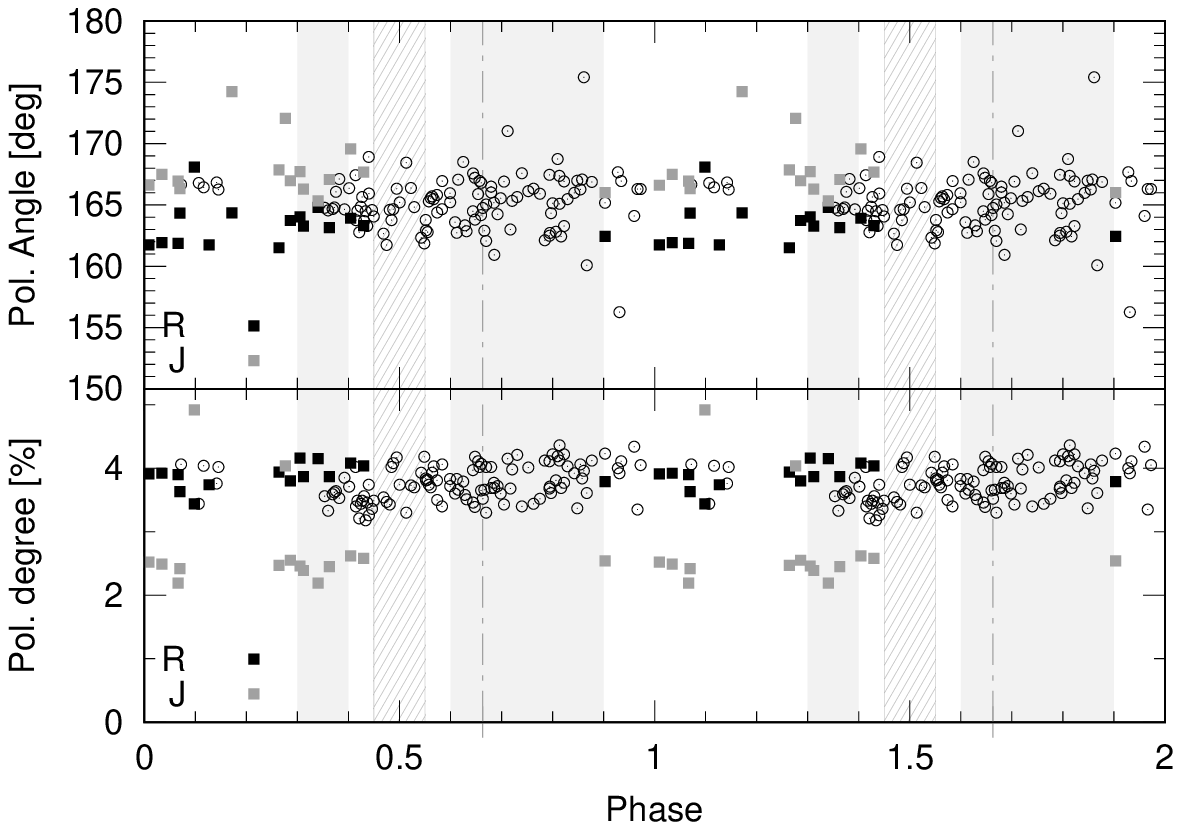}
\end{tabular}
\caption{The same as Figure \ref{fig:lsiPol}, but for \hess \ with Kanata/HOWPol (until MJD 57,042, $R$-band) and Kanata/HONIR (from MJD 57,330, $R$-band, and from MJD 57,139, $J$-band). 
See \mbox{Figure \ref{fig:hessPhotoIRSF_multip}} for the shaded and striped area.
\label{fig:hessPol}}
\end{figure}   


\section{Discussions}\label{sec:discussions}
Multi-year monitoring of the gamma-ray binaries in the optical and near-infrared bands was considered to cover enough orbital cycles and phases to enable comparison of the orbital variability among these systems.
The observed light curves showed a variety of features which differed from system to system, as well as from wavelength band to band.
Table \ref{tab:sumVariation} summarizes the variations in the three gamma-ray binaries.
\vspace{-10pt}

$ $\begin{specialtable}[H] 
\caption{Summary of the observed variations and the system parameters in the three gamma-ray binaries with a Be star.
\label{tab:sumVariation}}
\begin{tabular}{rp{0.16\textwidth}p{0.17\textwidth}p{0.18\textwidth}}
\toprule
	& \textbf{\psrb}	& \textbf{\hess} &\textbf{\lsi}  \\
\midrule
\multicolumn{4}{l}{System parameters} \\
\midrule
$P_\mathrm{orb}$ [days]		&  1236.7 	& 313--321  &   26.496 \\
$e$		&  0.87		& 0.64--0.83  &  0.537   \\
$R_\mathrm{disk}$(H$\alpha$) [$R_\odot$]		& 432--644 & 156--226  &  31--36 \\
$R_\mathrm{disk}$(H$\beta$) [$R_\odot$]		&  	202 & 52--66  &  18--19 \\
\midrule
\multicolumn{4}{l}{Orbital variability} \\
\midrule
Photometry & Near-infrared Brightening at periastron & Decrease at phase 0.3? & Sinusoidal variation \\
Polarization & No? & Sinusoidal variation? & Periodic variation \citep{Kravtsov2020} $^3$ \\
vs Emission lines & EW increases at the same timing \citep{Kawachi2021} & EW(H$\beta$) decreased when $V$-band flux decreased? & EW shows anti-correlation \citep{Zamanov2014} $^2$ \\
vs X/gamma-rays & Outbursts at periastron $^1$ \citep{Kawachi2021} & Anti-correlation with the primary outburst? & Anti-correlation \citep{Zamanov2014} \\
\bottomrule
\end{tabular}
\footnotesize \noindent{Notes: References for the orbital parameters; \psrb \ \cite{Shannon2014} ; \hess \ \cite{Bongiorno2011,Casares2012,Moritani2018}; \lsi \ \cite{Aragona2009}; References of the Be disk size; \psrb \ (see text for details): H$\alpha$ \citep{Chernyakova2020,Zamanov2016}, H$\beta$ \citep{Negueruela2011} ; \hess \ and \lsi \ \ \cite{Zamanov2016}; (1): Outbursts occurs at periastron, but the peak is at a different time; (2): Only blue part of double-peaked H$\alpha$  line profile; (3): 13-day periodicity was reported.} 
\end{specialtable}

In the gamma-ray binaries, the fluxes in the optical and near-infrared bands were dominated by that of the massive star.
Be disks exhibit emission-line components in Hydrogen (H I), Helium (He I), and other metallic lines (e.g., Fe II, Si II) in their spectra, whose profiles represent a Keplerian rotation disk.
Free--free and free--bound emissions from the Be disk can lead to excess in the near-infrared continuum. Disk absorption  affects the observed spectral energy distribution depending on viewing angle of the system \citep{Rivinius2013}.

The Be disk---isolated or in a binary system---changes its structure over various time scales, causing  variations in the observed magnitude, color, spectra, and polarization.
When the equatorial slow outflow from the photosphere stops, the Be disk dissipates and  
 the emission-line components, as well as the IR excess, disappear. 
Such variability is on a timescale from years to decades, much longer than the orbital period of the gamma-ray binaries, and likely occurs independently of the interaction with the compact object; and, hence, independently of the orbital phase.
In addition to the variability due to stellar activities, the existence of the compact object gives extra sources of perturbation to the Be disk in a binary system, through the tidal force and/or the plasma wind.
Due to the large, eccentric orbit, such a perturbation is expected to occur at a dedicated phase, thus resulting in orbital modulation.

In spectroscopy, line-profile variability indicates the structural changes of the Be disk such as density perturbation, precession, and warping.
In several Be stars `S-shaped' variation was observed; an emission component has moved blue-ward and red-ward in the line profile.
Moreover, many Be stars have shown `V/R oscillation'; relative intensity of the blue peak to the red peak of double-peaked profile has oscillated. 
These variabilities are thought to be caused by the one-armed oscillation \citep{Okazaki1991,Papaloizou1992,Aragona2010,Rivinius2013}.
Episodic change between the single- and double-peaked profiles in H$\alpha$ line has been observed in the Be/X-ray binary 4U~0115$+$63, implying the precession of the Be disk \citep{Reig2007}. 
A triple-peaked profile---indicating a precessing, warped Be disk---has been observed in the Be/X-ray binary A~0535$+$262 \citep{Moritani2013}.

In photometry, 
disk growth causes brightening of the magnitude in the optical and near-infrared bands for most Be stars. 
 The disk absorbs and scatters part of the stellar flux and if the disk  
  is seen close to edge-on, its growth may result in a net dimming of the system.
Since the radius of the emission region of the individual band differs \citep{Rivinius2013}, the color changes as the disk size changes. 
Indeed, EW($H\alpha$) and color ($J-Ks$) are correlated, as both observables represent the disk size \citep{Reig2011}.
Our observation showed that the average brightness in the optical and near-infrared bands did not change over several years.
This suggests that the overall size of the Be disk was stable over the observational period.

Linear polarization in the optical and near-infrared bands originates from the scattering of free electrons in the Be disk.
The polarization degree is roughly anti-correlated with the opacity, and the polarization angle is thought to be perpendicular to the major axis of the Be disk \citep{Rivinius2013}.
Statistical analysis of 672 Be stars has shown that the intrinsic polarization degree  is $<$1.5\% \citep{Yudin2001}.
Simulated polarization in the continuum band of the Be disk has predicted that the polarization parameters oscillate when the Be disk has precession and/or an azimuthally asymmetric structure \citep{Halonen2013a}.
It has also been shown through simulation that the polarization degree becomes saturated as the Be disk grows \citep{Halonen2013b}.
Spectropolarimetric observations of Be stars over decades have shown variation in $V$-band polarization as the Be disk dissipates and a new disk shows up \citep{Wisniewski2010}.
Polarimetric observation of the Be/X-ray binary 4U~0115$+$63, covering a giant outburst, has shown that the polarization degree and angle changed significantly while the Be disk warped \citep{Reig2018}.

\subsection{Orbital Modulation}
At the periastron passage, \psrb \ showed brightening of the near-infrared bands, indicating structural change of the Be disk due to tidal force from the pulsar~\citep{Kawachi2021}.
van Soelen et al. (2016) has reported an increase of EW (H$\alpha$) around the periastron, suggesting expansion or elongation of the Be disk \citep{vanSolen2016}.
Furthermore, the double-peaked profile of He I line from their observation suggested a hint of oscillation in the peak separation and V/R ratio (i.e., the ratio of the peak intensity of the blue part to that of the red part), indicating excitation and propagation of the density wave in the Be disk by tidal force from the pulsar.
These variations occurred at a similar time---about 10 days before periastron---and 
 showed a peak around 15 days after periastron.
This implies that the timescale of the tidal force impact on the Be disk of this system is around $\sim$25 days.

Similar to \psrb , \lsi \ also showed a possible brightening in the $V$ and $R$ bands at 
 the post-periastron passage (Figures \ref{fig:lsiPhoto_howpol} and \ref{fig:lsiPhoto_honir}), while it was not seen in \hess \ at either proposed periastron (Figures \ref{fig:hessPhotoIRSF}, \ref{fig:hessPhoto_howpol} and \ref{fig:hessPhoto_honir}).
However, S-shaped variation in the Balmer lines has been reported in \hess \ around \mbox{$\phi$$\sim$0.5--1.0 \citep{Aragona2010,Moritani2015}}.
If the periastron is at $\phi = 0.663$ \citep{Moritani2018}, the line-profile variability suggests excitation of the density wave.
If this were the case, brightening may be observed if precise photometric monitoring were carried out for this system.
Considering the timescale of the disk reconstruction seen in \psrb \ ($\sim$25 days for $P_\mathrm{orb} = 1236.72$ days, corresponding to the phase duration of $\sim$0.02), monitoring with a high cadence around periastron is needed to further discussion of this possibility.

In contrast to \psrb , \hess \ showed no turning point in color change (Figure \ref{fig:hessPhotoIRSF_h_jh}; as  the $J$-band brightness decreases, the $J-H$ color becomes reddened). 
The tendency of color-magnitude variation may be similar to   ``the negative correlation'' classified by  \citet{Harmanec1983} for shell stars; the stronger H I line emission, the fainter the star. According to \citet{Harmanec1983}, the edge-on Be stars should also grow redder as the star dims due to obscuration by  a growing dense disk. 
 The Balmer lines of \hess \ do not exhibit a shell profile \citep{Moritani2015} but the inclination angle {\textit i} of  $\simeq$47$^\circ$--80$^\circ$ has been estimated  \citep{Casares2012}. Our result may suggest a variation in size and/or density of the highly inclined disk to the line of sight.
In spectroscopy, \lsi \ \citep{Zamanov2014} and \hess \ (this work) did not show EW change at a dedicated phase as \psrb \ did; EW variation looked rather sinusoidal within one orbital cycle.

Orbital variations in polarization suggests structure changes in the Be disk, or an additional source of polarization or depolarization at a certain phase.
The lack of change in the polarization degree within 1 \%  in \psrb \, despite the variability in photometry and spectroscopy at periastron (Figure \ref{fig:pstbpol}), suggested that the Be disk is larger than \mbox{10 $R_*$ \citep{Halonen2013b}}.
Although our monitoring did not detect polarimetric variation in \lsi, another long-term monitoring study has shown such variation with a period of half of the orbital period \citep{Kravtsov2020}.
A small change ($\sim$0.5\%) in polarization degree in \hess \ seemed to occur on a timescale on the order of several tens of days to one orbit (Figures \ref{fig:hessPol}), thus not being a short event at a dedicated phase ($\sim$10 days), such as the periastron event of \psrb .
These features suggest that the disk structure changes in this timescale; i.e., changing and recovering the structure while the compact object revolves around the orbit.
Therefore, the tidal force is rather unlikely to be the origin of such polarimetric variation.

\subsection{Long-Term Be Disk Activity}

As described above, the Be disk itself changes structure due to stellar activities.
The timescale of Be activity is longer than the orbital period and is thought to cause super-orbital variation, as seen in \lsi .
Our monitoring observation did not show a sign of super-orbital modulation for the other systems, although the coverage of orbital cycles and/or accuracy of the measurement are not sufficient to determine this conclusively.
Please note that the H$\alpha$ emission line of \hess \ strengthened in 2018---after our monitoring period---thus indicating Be disk growth \citep{Stoyanov2018}.

\subsection{Comparison with High-Energy Emissions and Orbital Parameters}
From the viewpoint of the relationship with X-ray and gamma-ray activities, the near-infrared brightness in \psrb \ increased, while high-energy emissions increased around periastron.
On the contrary, \lsi \ showed anti-correlation between the optical brightness and high-energy emissions, while \hess \ showed anti-correlation at the primary outburst.
The color changed monotonously as the brightness increased and decreased in  \hess \ (Figure \ref{fig:hessPhotoIRSF_h_jh}), in contrast to what is expected from reconstruction by the tidal force, as seen in \psrb \ \citep{Kawachi2021}.
This feature suggests that the size of the Be disk had simply changed, due to either mass transfer or stripping by the pulsar wind.

As shown in Table \ref{tab:sumVariation}, the orbit size was quite different among the three considered gamma-ray binaries.
The Be disk is larger with a larger orbit, as the Be disk is truncated by the compact object.
Here, we estimated the Be disk size in \psrb \ using the relationship between EW and emission region of H$\alpha$ (\citep{Zamanov2016} eq. 8), citing EW at the 2010 periastron \citep{Chernyakova2014}.
We also estimated the Be disk size by the emitting region of H$\beta$ of \psrb , from the peak separation of the double-peaked profile (\citep{Negueruela2011} $\sim$$116~\mathrm{km \; s^{-1}}$) assuming  a Keplerian disk.
The disk size of \lsi \ and \hess \ were cited from the previous study \citep{Zamanov2016}.
The observed variation was also different, which seems to be related to the differences in orbital parameter and nature of the system. 
The light curve showed a clear sinusoidal pattern in \lsi , the binary with the smallest orbit, where the Be disk fills a large fraction of area of the orbit compared with the other systems.
On the other hand, \psrb , where the Be disk covers only a small area of the orbit, showed variations only at periastron.
\hess \ shows  similar variations with both \psrb \ and \lsi .
Compared to \psrb, \hess \ has \mbox{$\sim$8 times} smaller orbit, whereas the Be disk is $\sim$3--4 times smaller.
On the other hand, compared to \lsi \, \hess \ has $\sim$40 times larger orbit, whereas the Be disk is $\sim$3--4 times larger.
These features suggest that \hess \ is in the middle of \psrb \ and \lsi \ from the viewpoint of the size if the Be disk in comparison to the orbit; although the alignment of the Be disk plane with respect to the orbital plane also affects the interactions.

Besides the Be disk size compared to the orbit, other parameters also have impacts on binary interaction, such as the orbit eccentricity, misalignment angle between the Be disk and orbit, and Be disk base density.
The impact of the disk truncation is weaker if the Be disk is misaligned to the orbit of the companion star, which enables the disk to become larger.
Smoothed particle hydrodynamics (SPH) simulation of the density wave in the Be disk in binary systems has shown that spiral arms and the radial density distribution depends on the misalignment angle between the Be disk and the orbital plane, as well as the  viscosity of the Be disk \citep{Cyr2020}.
In a Be disk with higher misalignment angle and higher viscosity, the spiral arm becomes more tight and radial density distribution becomes flatter, without a clear cut-off.

We monitored several Be/X binaries, which have similar orbit size and eccentricity to the gamma-ray binaries but have accreting pulsars.
Line-profile variabilities implying the density wave have also been observed in several Be/X-ray binaries.
Comparison of orbital modulation between these systems may provide clues to discuss the relationships between these systems.






\section{Concluding Remarks}\label{sec:conclusion}

Three gamma-ray binaries hosting a Be star were monitored, in terms of their optical and near-infrared brightness and linear polarization, from 2010 to 20108.
The long-term light curves indicated the orbital modulations (or their sign) in all the systems, although the variations differed from system to system.
All three binaries showed variation due to the tidal force from the compact object, although more precise monitoring with a higher cadence is required to determine more conclusive results.
The diversity of the variation is possibly related to the difference in the size and base density of the Be disk, the disk size compared to the orbit, and the alignment of the Be disk plane with respect to the orbital~plane.



\vspace{6pt} 



\authorcontributions{Conceptualization, Y.M. and A.K.; investigation and formal analysis, Y.M. and A.K.; data curation, Y.M. and A.K.; visualization, Y.M. and A.K.; writing---original draft preparation, Y.M.; writing---review and editing, Y.M. and A.K. All authors have read and agreed to the published version of the manuscript.}

\funding{We acknowledge the support by the Japan Society for the Promotion of Science (JSPS) KAKENHI Grant Numbers JP21540304, and Research Fellowships for the Promotion of Science for Young Scientists.
This work was supported as a Joint Research Project under agreement between JSPS and the National Research Foundation (NRF) of South Africa. The IRSF project is a collaboration between Nagoya University and the South African Astronomical Observatory (SAAO) supported by the Grants-in-Aid for Scientific Research on Priority Areas (A) (No. 10147207 and No. 10147214), Optical \& Near-Infrared Astronomy Inter-University Cooperation Program, the Ministry of Education, Culture, Sports, Science and Technology (MEXT) of Japan and NRF of South Africa.
}

\institutionalreview{Not applicable}

\informedconsent{Not applicable}

\dataavailability{IRSF images are not stored on public archive, so one should contact the observatory to request the data. Kanata images of this work are archived on SMOKA science data archive system run by Astronomical Data Archives Center (ADAC), Astronomy Data Center (ADC), National Astronomical Observatory of Japan (NAOJ) (\url{https://smoka.nao.ac.jp/).} The data is public 18-month after observation. One can search for and request the data. }

\acknowledgments{The authors are grateful to S.~Chimasu of Tokai University for his contribution in the observations and the early-stage analysis of \hess \ data.
}

\conflictsofinterest{The authors declare no conflict of interest.} 



\abbreviations{Abbreviations}{
The following abbreviations are used in this manuscript:\\

\noindent 
\begin{tabular}{@{}ll}
HMXB & High-Mass X-ray Binary \\
IRSF &  InfraRed Survey Facility \\
SIRIUS & Simultaneous InfraRed Imager for Unbiased Survey \\
SIRPOL & SIRius POLarimetry mode \\
HOWPol & Hiroshima One-shot Wide-field Polarimeter \\
HONIR & Hiroshima Optical and Near-InfraRed camera \\
EW & Equivalent Width
\end{tabular}}

\appendixtitles{yes} 
\appendixstart
\appendix
\section{Relative (Differential) Magnitude with IRSF and Kanata}

To compare with IRSF data, we derived differential magnitude for the Kanata data.
The differential magnitude of IRSF/SIRIUS data is described as follows:

\begin{equation}\label{eq:irsfmag}
    \Delta m_\mathrm{IRSF} = m_\mathrm{t}(t) - m_\mathrm{t}(0) - \frac{\Sigma w_i( m_{\mathrm{r},i}(t) - m_{\mathrm{r},i}(0) )  )}{\Sigma w_i} 
\end{equation}

\begin{equation}
    \sigma _{\Delta m_\mathrm{IRSF}} = \sqrt{ \frac{<m_{\mathrm{r},i}(t) - m_{\mathrm{r},i}(0)>^2}{\sqrt{n}} + \sigma ^2_{m_\mathrm{t}(t)} }
\end{equation}

Here, $m(t)$ is magnitude from the aperture photometry at given time $t$, and $m(0)$ is that at a reference time ($t=0)$.
The last term indicates the weighted mean of the comparison stars with the weight of $w_i$, which is related to IRAF photometry error.

Assuming the weights are uniform ($w_i=1$) among the comparison stars, the \mbox{Equation \ref{eq:irsfmag}} becomes

\begin{align}
    \Delta m_\mathrm{IRSF} &= m_\mathrm{t}(t) - m_\mathrm{t}(0) - \frac{\Sigma ( m_{\mathrm{r},i}(t) - m_{\mathrm{r},i}(0) )  )}{N} \\
    &= \frac{\Sigma ( m_{\mathrm{t}}(t) - m_{\mathrm{r},i}(0) ) -  \Sigma ( m_{\mathrm{t}}(0) - m_{\mathrm{r},i}(0) ))}{N}\\
    &= \Delta m_\mathrm{Kanata}
\end{align}\label{eq:irsfmag2}

In this case, error in $\Delta m_\mathrm{Kanata}$ should be RMS of each measurement.



\end{paracol}

\printendnotes[custom]

\reftitle{References}

\end{document}